\documentclass[twocolumn,published]{aastex701}

\newcommand{\Sec}[1]{Section~\ref{#1}}

\newcommand{\Eq}[1]{Equation~\ref{#1}}
\newcommand{\Fig}[1]{Figure~\ref{#1}}
\newcommand{\Tab}[1]{Table~\ref{#1}}

\usepackage{graphicx}	
\usepackage{float}
\usepackage{subcaption}
\usepackage{wrapfig}
\usepackage[autostyle]{csquotes}
\usepackage{amsmath}	
\usepackage{tikz}
\bibpunct[, ]{(}{)}{,}{a}{,}{,}
\usetikzlibrary{arrows.meta,shapes, positioning, decorations.pathmorphing, arrows.meta, calc, shadings, positioning, fadings, shadows}
\journalinfo{arXiv preprint}

\begin{document}

\title{SAGE26 Paper I: Modelling the baryon cycle from cosmic dawn to the
present day}

\author[orcid=0009-0005-8465-2472]{Michael Bradley}
\affiliation{Centre for Astrophysics and Supercomputing, Swinburne University of Technology, PO Box 218, Hawthorn VIC 3122, Australia}
\affiliation{ARC Centre of Excellence for Dark Matter Particle Physics (CDM)}
\email[show]{mbradley@swin.edu.au}

\author[orcid=0000-0002-5009-512X]{Darren J. Croton}
\affiliation{Centre for Astrophysics and Supercomputing, Swinburne University of Technology, PO Box 218, Hawthorn VIC 3122, Australia}
\affiliation{ARC Centre of Excellence for Dark Matter Particle Physics (CDM)}
\affiliation{ARC Centre of Excellence for All Sky Astrophysics in 3 Dimensions (ASTRO 3D)}
\email[show]{dcroton@swin.edu.au}

\author[orcid=0000-0002-5425-0641]{Robert Adriel Mostoghiu Paun}
\affiliation{Centre for Astrophysics and Supercomputing, Swinburne University of Technology, PO Box 218, Hawthorn VIC 3122, Australia}
\affiliation{ARC Centre of Excellence for Dark Matter Particle Physics (CDM)}
\affiliation{ARC Centre of Excellence for All Sky Astrophysics in 3 Dimensions (ASTRO 3D)}
\email{rmostoghiu@swin.edu.au}

\author[orcid=0000-0003-0250-3827]{Dhruba Dutta Chowdhury}
\affiliation{Centre for Astrophysics and Supercomputing, Swinburne University of Technology, PO Box 218, Hawthorn VIC 3122, Australia}
\affiliation{ARC Centre of Excellence for Dark Matter Particle Physics (CDM)}
\email{dhruba.duttachowdhury@mail.huji.ac.il}

\author[orcid=0009-0001-5653-9481]{Jayde Willingham}
\affiliation{Centre for Astrophysics and Supercomputing, Swinburne University of Technology, PO Box 218, Hawthorn VIC 3122, Australia}
\affiliation{ARC Centre of Excellence for Dark Matter Particle Physics (CDM)}
\email{jwillingham@swin.edu.au}

\begin{abstract}
We present SAGE26, a comprehensive update to the Semi-Analytic Galaxy Evolution (SAGE) model. SAGE26 integrates five observationally motivated improvements: (1) distinct circumgalactic medium and hot-halo gas physics, (2) molecular hydrogen for star formation, (3) velocity- and redshift-dependent supernova feedback, (4) bulge radii tracking through time, and (5) feedback-free burst galaxies. Each improvement works to regulate galaxy formation from cosmic dawn to the present. Most significantly, the addition of feedback-free burst galaxies provides a pathway to produce high-redshift massive galaxies at the abundances observed by the James Webb Space Telescope. Furthermore, the new circumgalactic medium framework improves the treatment of baryons by distinguishing between free-fall-limited inflow in low-mass haloes and hydrostatic cooling in massive groups and clusters using a characteristic halo shock mass. SAGE26 reproduces key observational constraints across cosmic time, including the stellar mass function and cosmic star formation rate density, while retaining the computational efficiency and flexibility of its predecessor. The code and data products are open-source and publicly available.
\end{abstract}


\section{Introduction}
Galaxy formation remains one of the most complex problems in astrophysics, requiring models that bridge scales from dark matter haloes to individual star-forming regions across cosmic time. Semi-analytic models (SAMs) have emerged as powerful tools for understanding this complexity, offering the computational efficiency needed to explore large cosmological volumes while maintaining sufficient physical detail to make testable predictions \citep{white1991,kauffmann1999,somerville1999}.\\
\\
The Semi-Analytic Galaxy Evolution model (SAGE; \citealt{croton2006,croton2016}), a legacy framework of galaxy formation theory for nearly two decades, provides crucial insights into the cosmic evolution of galaxies in cosmological simulations. The 2006 version introduced the widely used radio and quasar modes for describing the behaviour of supermassive black holes and their effects on the baryon cycle in dark matter haloes. The 2016 version of SAGE (SAGE16) was designed to run on any simulation, be highly computationally efficient, and produce accurate predictions at $\rm z=0$ for numerous galactic properties. However, as new observations have emerged and new physical processes have been identified, the need for a comprehensive modernisation of SAGE's underlying physics has become apparent.\\
\\
The shift in focus beyond the local Universe has exposed limitations in many popular galaxy formation models \citep{boylan-kolchin2023,lagos2024,lu2024,boylan-kolchin2025,fontanot2026}, including SAGE. While SAGE16 remains a robust tool for the local Universe, it was not built to explore the parameter space opened up by modern, high- and ultra-high-redshift surveys. Some overarching problems have emerged, such as the overproduction of dwarf-quiescent galaxies at many epochs, and the underproduction of massive galaxies at early epochs. The latter has become an especially interesting area of research, as recent observations have revealed an abundance of massive bright galaxies at $\rm z\geq6$ \citep{donnan2022,labbe2023,finkelstein2023,harikane2023}.\\
\\
To address these questions, we present SAGE26 in this paper. Rather than tuning parameters to suppress overproduction or boost galaxy growth, SAGE26 integrates new physical recipes that work together to regulate the baryon cycle. We update the model to distinguish between smaller systems that interact with a circumgalactic medium and more massive systems that develop cold streams penetrating through hot haloes at early epochs, and implement a feedback-free burst mode to capture the highly efficient star formation at early times of the Universe. These changes, coupled with an updated star formation recipe and supernova feedback prescription, better match how we observe galaxies evolving.
\section{An overview of SAGE26}\label{SAGE26_overview}
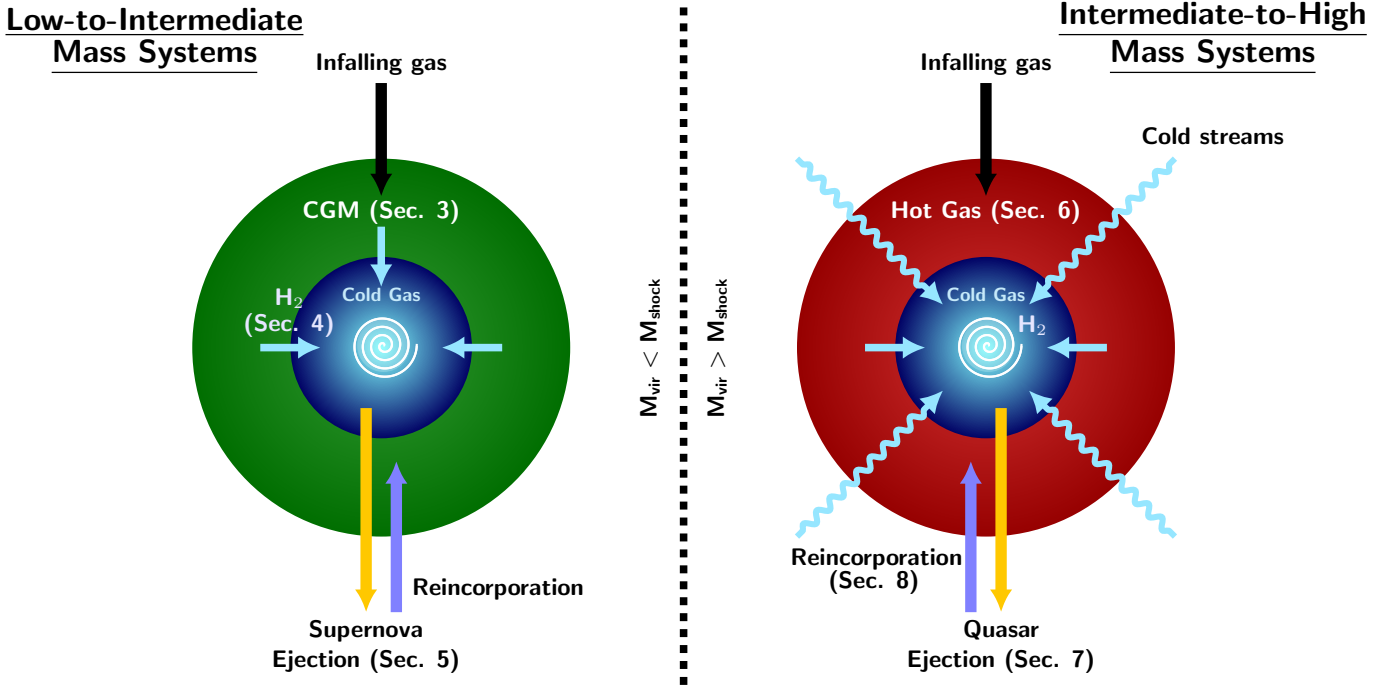
\begin{figure*}
    \definecolor{sageRed}{RGB}{200, 0, 0}
    \definecolor{sageDarkRed}{RGB}{150, 0, 0}
    \definecolor{sageGreen}{RGB}{0, 128, 0}
    \definecolor{sageDarkGreen}{RGB}{0, 110, 0}
    \definecolor{sageBlue}{RGB}{0, 100, 255}
    \definecolor{sageDarkBlue}{RGB}{0, 0, 100}
    \definecolor{sageYellow}{RGB}{255, 200, 0}
    \definecolor{sageLightBlue}{RGB}{150, 230, 255} 
    
    \begin{tikzpicture}[
        font=\sffamily\bfseries,
        >={Latex[width=3mm,length=3mm]},
        arrow line/.style={draw, line width=1.5mm},
        label text/.style={fill=none, align=center},
        cooling arrow/.style={
            ->, draw=sageLightBlue, line width=0.8mm,
            decorate,
            decoration={snake, amplitude=0.6mm, segment length=2.5mm,
                        pre length=1mm, post length=2.5mm}
        },
        galaxy/.style={
            execute at begin node={
                \begin{tikzpicture}[scale=0.35]
                    \fill[white, opacity=0.2] (0,0) circle (1);
                    \draw[white, thick, domain=0:25.13, variable=\t, samples=200, smooth]
                        plot ({\t r}: {0.05*\t});
                    \draw[white, thick, domain=0:25.13, variable=\t, samples=200, smooth]
                        plot ({\t r + 3.14}: {0.05*\t});
                \end{tikzpicture}
            }
        }
    ];

    \node[anchor=south] at (-7, 4.0) {\large \underline{\textbf{Low-to-Intermediate}}};
    \node[anchor=south] at (-7, 3.5) {\large \underline{\textbf{Mass Systems}}};
    \node[anchor=south] at (7, 4.0) {\large \underline{\textbf{Intermediate-to-High}}};
    \node[anchor=south] at (7, 3.5) {\large \underline{\textbf{Mass Systems}}};
    
    
    \shade[inner color=sageGreen!80!white, outer color=sageDarkGreen] (-4,0) circle (2.5cm);
    \node[text=white, yshift=1.8cm] at (-4,0) {CGM (\textit{Sec. 3})};
    
    \shade[inner color=cyan!50!white, outer color=sageDarkBlue] (-4,0) circle (1.2cm);
    \node[text=cyan!20!white, yshift=0.7cm, scale=0.8] at (-4,0) {Cold Gas};
    
    \node[galaxy] at (-4,0) {};
    \node[text=blue!10!white, left=0.3cm] at (-4.6,0.65) {H$_2$};
    \node[text=blue!10!white, left=0.3cm] at (-4.2,0.3) {(\textit{Sec. 4})};
    
    
    \draw[->, line width=1.5mm, black] (-4, 3.5) -- (-4, 2.0);
    \node[above] at (-4, 3.5) {Infalling gas};
    
    \draw[->, line width=1mm, sageLightBlue] (-4, 1.6) -- (-4, 0.8);
    \draw[->, line width=1mm, sageLightBlue] (-5.6, 0) -- (-4.8, 0);
    \draw[->, line width=1mm, sageLightBlue] (-2.4, 0) -- (-3.2, 0);
    
    \draw[->, arrow line, sageYellow] (-4.2, -0.8) -- (-4.2, -3.5);
    \node[below, align=center, text=black] at (-4.2, -3.5) {Supernova\\Ejection (\textit{Sec. 5})};
    
    \draw[->, arrow line, blue!50!white] (-3.8, -3.5) -- (-3.8, -1.5);
    \node[left, text=black] at (-1.2, -3.2) {Reincorporation};

    
    \draw[dashed, line width=1mm, black] (0, 4.5) -- (0, -4.5);
    
    \node[rotate=90, anchor=south] at (-0.2, 0) {M$_{\text{vir}} < \text{M}_{\text{shock}}$};
    
    \node[rotate=90, anchor=north] at (0.2, 0) {M$_{\text{vir}} > \text{M}_{\text{shock}}$};


    
    \shade[inner color=sageRed!80!white, outer color=sageDarkRed] (4,0) circle (2.5cm);
    \node[text=white, yshift=1.8cm] at (4,0) {Hot Gas (\textit{Sec. 6})};
    
    \shade[inner color=cyan!50!white, outer color=sageDarkBlue] (4,0) circle (1.2cm);
    \node[text=cyan!20!white, yshift=0.7cm, scale=0.8] at (4,0) {Cold Gas};
    
    \node[galaxy] at (4,0) {};
    \node[text=blue!10!white, right=0.3cm] at (4,0.3) {H$_2$};

    \node[text=black, yshift=1.8cm] at (7,1) {Cold streams};
    
    
    \draw[->, line width=1.5mm, black] (4, 3.5) -- (4, 2.0);
    \node[above] at (4, 3.5) {Infalling gas};
    
    \draw[->, line width=1mm, sageLightBlue] (2.4, 0) -- (3.2, 0);
    \draw[->, line width=1mm, sageLightBlue] (5.6, 0) -- (4.8, 0);
    \draw[cooling arrow] (1.500,  2.500) -- (3.434,  0.566);
    \draw[cooling arrow] (6.500,  2.500) -- (4.566,  0.566);
    \draw[cooling arrow] (1.500, -2.500) -- (3.434, -0.566);
    \draw[cooling arrow] (6.500, -2.500) -- (4.566, -0.566);
    
    \draw[->, arrow line, sageYellow] (4.2, -0.8) -- (4.2, -3.5);
    \node[below, align=center, text=black] at (4.2, -3.5) {Quasar\\Ejection (\textit{Sec. 7})};
    
    \draw[->, arrow line, blue!50!white] (3.8, -3.5) -- (3.8, -1.5);
    \node[right, text=black] at (1.3, -2.8) {Reincorporation};
    \node[right, text=black] at (1.8, -3.1) {(\textit{Sec. 8})};
    
    \end{tikzpicture}
    \caption{The new baryon cycle schematic for the circumgalactic medium and hot halo, distinguished by the characteristic halo mass $\rm M_{shock}=6\times10^{11}\ M_{\odot}$. The left side describes the galactic fountain for haloes $\rm M_{vir}<M_{shock}$, where haloes interact with a circumgalactic medium and gas flows directly onto the galaxy. The right side shows the more traditional cooling and heating baryon cycle for haloes $\rm M_{vir}>M_{shock}$, where little has changed from the SAGE16 recipe except the addition of cold streams (wiggly arrows).}
    \label{fig:SAGE26_schematic}
\end{figure*}
\begin{table*}
\centering
\begin{tabular}{lccccccccc}
\hline
\hline
Simulation & $\rm N_{\text{part}}$ & $\rm M_{\text{part}}$ & $\rm L_{\text{box}}$ & $\rm \Omega_M$ & $\sigma_8$ & h & Code & Subhalo finder & Tree constructor \\
 & & $\rm [h^{-1}\ M_{\odot}]$ & [$\rm h^{-1}$\ cMpc] & & & & & \\
\hline
Millennium & $2160^3$ & $8.60 \times 10^8$ & 500 & 0.250 & 0.900 & 0.73 & \textsc{gadget-2} & \textsc{subfind} & \textsc{l-halotree}\\
miniUchuu & $2560^3$ & $3.27 \times 10^8$ & 400 & 0.308 & 0.815 & 0.67 & \textsc{gadget-2} & \textsc{rockstar} & \textsc{consistent-trees}\\
\hline
\hline
\end{tabular}
\caption{Details of the N-body simulations used in the analysis of SAGE26 in this paper. The columns provide particle number, $\rm N_{part}$; particle mass, $\rm M_{part}$; periodic box length, $\rm L_{box}$ in comoving units; dimensionless cosmological matter density, $\rm \Omega_M$ (for which the equivalent for dark energy is $\rm \Omega_{\Lambda}=1-\Omega_{M}$); and the root-mean-square mass fluctuation within a sphere of radius $\rm 8h^{-1}\ Mpc$, $\sigma_8$.}
\label{tab:simulation_details}
\end{table*}
\begin{table*}
\footnotesize
\begin{tabular}{lllll}
\hline
\hline
Parameter                & Description                                                            & SAGE16           & SAGE26          & Fixed \\
\hline
\hline
$f_{\rm b}^{\rm cosmic}$ & (Cosmic) baryon fraction                                               & 0.17, 0.15    & 0.17, 0.15          & Yes    \\
\hline
$\rm z_0$                & Redshift when H II regions overlap                                      & 8.0           & 8.0                 & Yes   \\
$\rm z_r$                & Redshift when the intergalactic medium is fully reionised              & 7.0           & 7.0                 & Yes   \\
\hline
$\rm \alpha_{SF}$        & Star formation efficiency                                              & 0.05          & 0.05                 & No    \\
$\rm \alpha_{eff}$       & FFB galaxies star formation efficiency                         & \nodata             & 0.20                 & No    \\
$Y$                      & Yield of metals from new stars                                        & 0.025         & 0.025                & Yes   \\
$R$                      & Instantaneous recycling fraction                                       & 0.43          & 0.43                 & Yes   \\
\hline
$\rm \epsilon_{disk}$    & Mass-loading factor due to supernovae/FIRE normalisation in SAGE26                                  & 3.0           & 2.9                  & No    \\
$\rm \epsilon_{halo}$    & Efficiency of supernovae to unbind gas from the hot halo               & 0.3           & 0.3                  & No    \\
$\rm \kappa_{reinc}$     & Velocity scale for gas reincorporation                            & 0.15          & 0.19                 & Yes   \\
\hline
$\rm \kappa_{R}$         & Radio mode feedback efficiency                                         & 0.08          & 0.08                 & No    \\
$\rm \kappa_{Q}$         & Quasar mode feedback efficiency                                        & 0.005         & 0.005                & No    \\
$f_{\rm BH}$             & Rate of black hole growth during quasar mode                           & 0.015         & 0.015                & No    \\
\hline
$f_{\rm friction}$       & Threshold subhalo-to-baryonic mass for satellite disruption or merging & 1.0           & 1.0                  & Yes   \\
$f_{\rm major}$          & Threshold mass ratio for merger to be major                            & 0.3           & 0.3                  & Yes   \\
\hline
$\rm \eta_{SN}$          & Number of supernovae per solar mass formed of stars                    & 0.005           & 0.005                  & Yes   \\
$\rm E_{SN}$             & Supernova energy in erg                                               & $10^{51}$       & $10^{51}$              & Yes   \\
\hline
\hline
\hline
\end{tabular}
\caption{Fiducial parameters for SAGE26 with Millennium or miniUchuu simulations. Each parameter has a description, the value used in SAGE16 and SAGE26, and whether they are fixed or a free parameter. For $f_{\rm b}^{\rm cosmic}$ in both SAGE16 and SAGE26, we use $f_{\rm b}^{\rm cosmic}=0.17$ for Millennium and $f_{\rm b}^{\rm cosmic}=0.15$ for miniUchuu.}
\label{tab:parameters}
\end{table*}
SAGE26 incorporates multiple physics improvements that draw inspiration from other semi-analytic models such as the Santa Cruz SAM \citep{somerville2015}, SHARK \citep{lagos2018,lagos2024}, GAEA \citep{delucia2024,fontanot2025}, and DarkSAGE \citep{stevens2024}, while adding novel physical recipes to SAGE16. Key changes that distinguish SAGE16 from SAGE26 are summarised here:
\begin{itemize}
    \item \textit{Circumgalactic Medium and Hot-Intracluster / Intragroup Regimes:} We implement a new framework to distinguish between haloes with a circumgalactic medium and haloes with a hot intracluster or intragroup medium. This results in more realistic cooling and gas inflow based on halo mass \citep{dekel2006,carr2023}.
    \item \textit{Atomic and Molecular Hydrogen:} The cold gas reservoir has been divided into atomic and molecular hydrogen. Most stellar mass now forms directly from the $\rm H_2$ gas reservoir rather than from the total cold gas reservoir. This new gas tracking decreases the number densities of low-mass quiescent galaxies over cosmic time \citep{blitz2006}.
    \item \textit{Stellar Feedback:} Supernovae can reheat and eject gas from their host galaxy, where the amount of mass reheated per unit mass of stars formed is called the mass-loading factor, denoted by $\rm \eta_{reheat}$. SAGE26 derives $\rm \eta_{reheat}$ using ejection calculations from the FIRE simulation \citep{muratov2015} that depend on redshift and depth of the potential well. This has consequences for the galactic fountain: contrary to SAGE16, where the mass-loading factor was constant, supernovae can now reheat and eject variable amounts of gas per unit mass of new stars formed, which correlate with halo mass and redshift.
    \item \textit{Improved morphology:} As an extension to the merging and disk instability recipe, SAGE26 now explicitly tracks the bulge radius through time. Following \cite{tonini2016}, we distinguish between merger-driven and instability-driven bulge formation and growth.
    \item \textit{Feedback-free Burst Galaxies:} We implement a model for feedback-free bursts in galaxies, able to reproduce galactic properties such as mass and star formation rate, as observed at very high redshifts, seen by the James Webb Space Telescope (JWST). Following \cite{dekel2023} and \cite{li2024}, and depending on their host halo mass and redshift, a subset of such galaxies exhibit high star formation efficiencies, resulting in a higher number density of massive, high-redshift galaxies.
\end{itemize}
One of the major changes in SAGE26 is the reframing and extension of the SAGE16 hot and cold accretion modes for the interaction of infalling gas and the halo itself. \Fig{fig:SAGE26_schematic} illustrates this new galactic fountain, where we divide the halo population into two distinct regimes. To differentiate the halo regimes, we use the mass at which a halo can sustain a virial shock that heats infalling gas, as outlined in \cite{dekel2006}. To determine which inflow mode the halo will experience, we use a characteristic shock mass, $\rm M_{shock}=6\times10^{11}\ M_{\odot}$, which we assume is independent of redshift. This threshold arises from the shock stability criterion: a stable virial shock can only be sustained in a halo when the post-shock compression rate is sufficient to restore the pressure against radiative cooling losses \citep{dekel2006}.\\
\\
Haloes above the shock mass ($\rm M_{vir} > M_{shock}$) typically follow the process outlined in SAGE16, where infalling gas is shock-heated to the dark matter halo's virial temperature. We now label this gas as the hot intragroup or intracluster medium. Gas can then slowly cool into the interstellar medium, where it can fuel star formation. Additionally, we include cold streams that penetrate the hot gas (see \Sec{Coldstreams_method}).\\
\\
The primary feedback mechanism regulating galaxy growth in such high-mass haloes comes from the supermassive black hole within them, which reheats cooling gas through the radio Active Galactic Nuclei (AGN) mode and ejects gas with quasar winds outside the halo (see \Sec{AGN_method}). If conditions are met, such gas can be reincorporated from the hot gaseous atmosphere at a later time.\\
\\
For haloes below $\rm M_{shock}$ ($\rm M_{vir} < M_{shock}$), we introduce changes to the galactic fountain, where such galaxies now form and interact with a circumgalactic medium (see \Sec{CGM_method}). Within these haloes, infalling gas is deposited into the circumgalactic medium, and it can flow quickly into the galaxy's interstellar medium. Growth in these galaxies is primarily regulated via supernova feedback, causing winds to reheat the surrounding interstellar medium and eject gas into the ejected mass reservoir (for updates to the supernova reheating and ejection model, see \Sec{SN_method}). If these galaxies grow to be massive, reincorporation can begin to replenish the halo gas (see \Sec{Reinc_method}).\\
\\
We calibrate this model using a custom particle swarm optimisation package (SAGEswarm), which provides a powerful method for observational-constraint-based parameter-space exploration. We follow \cite{lagos2018,cora2018} and \cite{stevens2024} in using the particle swarm optimisation method, which is more computationally efficient and well-suited to degenerate, complex, or large parameter spaces \citep{ruiz2015}. We use the stellar mass function, the $\rm H_2$ mass function, the black hole-bulge mass relation and the cosmic star formation rate history density as constraints in parallel with visual inspection and comparison (see \Sec{calibration}).\\
\\
For this work, the model is run on top of the Millennium simulation \citep{springel2005a} with a volume of $\rm (500\ h^{-1}\ Mpc)^{3}$; and the miniUchuu \citep{ishiyama2021} simulation, with a volume of $\rm (400\ \ h^{-1}\ Mpc)^{3}$, which is a smaller volume than Millennium but has a higher mass resolution (see \Tab{tab:simulation_details}). The choice of these simulations is twofold: firstly, for continuity with SAGE16, which was calibrated on Millennium. Secondly, SAGE16 yielded inconsistent results between the two simulations, which needed to be resolved. Where relevant, we also use a \cite{chabrier2003} initial mass function to produce stellar masses. We provide a table of parameters for the fiducial model, with the values used for the SAGE26 model, derived from SAGEswarm and referenced throughout this paper (\Tab{tab:parameters}).\\
\\
In the following sections, we explore the more technical aspects of the SAGE26 update and describe the methodological details for this physics implementation. We begin with low-to-intermediate mass haloes that experience an altered galactic fountain. We also discuss the star formation and feedback processes responsible for galaxy growth in these systems. We then move into the massive systems, where more traditional galactic fountain physics dominates, but with the additional accretion pathway of cold streams at high and ultra-high redshifts.
\section{The gas cycle: Low-to-intermediate mass systems}
As part of restructuring the baryon cycle in SAGE, and with the introduction of the circumgalactic medium, we implement a new inflow mechanism to approximate its complex multi-phase physics.\\
\\
SAGE26 follows SAGE16, where the total baryonic content (stars, hot gas, cold gas, ejected gas, etc.) of each halo is maintained at the universal baryon fraction, $f_{\rm b}$. The mass difference, $f_{\rm b}\ \rm M_{vir}-m_b$ (where $\rm m_b$ is the total mass of baryons present in the previous timestep), is added at each timestep as pristine infalling gas brought in by the growing halo.\\
\\
The epoch of reionisation follows SAGE16, where the effect of photoionisation heating is modelled by defining a filtering mass below which the baryonic fraction in the halo is reduced relative to the universal value. To approximate the filtering mass, \cite{kravtsov2004} fitted an analytic model, defining two parameters to characterise the transition, $\rm z_0$ which marks the redshift where the first H II regions overlap, and $\rm z_r$ which is when the intergalactic medium is fully reionised (see \citealt{croton2016} for more details).\\
\\
For haloes below the shock mass ($\rm M_{vir} < M_{shock}$), we implement an altered gas physics framework from SAGE16. In these systems, this infalling pristine gas is deposited into the circumgalactic medium without a stable virial shock forming at the virial radius. The gas can then flow into the galaxy.
\subsection{Circumgalactic medium and the galactic fountain}\label{CGM_method}
The cooling time for a parcel of CGM gas is approximated by the ratio of its thermal energy to the cooling rate:
\begin{equation}\label{eq:cgmcooling}
    \rm t_{\rm cool}=\frac{3}{2}\, \frac{\rm \bar{\mu}\, \rm m_{\rm p}\, \rm k\, \rm T_{\rm vir}}{\rm \rho_{\rm CGM}\, \rm \Lambda(\rm T_{\rm vir}, \rm Z)}.
\end{equation}
Here, $\rm k$ is the Boltzmann constant, $\rm T_{vir}$ is the virial temperature of the halo and is recomputed at each timestep. For this regime of haloes in SAGE26, $\rm T_{vir}$ spans $\rm \sim3\times10^4-5\times10^5\ K$, hydrogen is therefore collisionally ionised across essentially the entire population, and $\rm \bar{\mu}=0.59$, which is the mean molecular weight for fully ionised gas, is self-consistent with the adopted temperature. $\rm m_p$ is the proton mass and $\rm \Lambda(T_{vir}, Z)$ is the cooling function \citep{croton2006} dependent on CGM gas temperature and metallicity.\\
\\
Following \cite{carr2023}, we model the CGM density, $\rm \rho_{CGM}$, as a spherically symmetric power-law profile, with slope 1.4 between $\rm 0.1R_{vir}$ and $\rm R_{vir}$, where $\rm R_{vir}$ is the halo's virial radius. We then substitute the mass-weighted average density from this profile into \Eq{eq:cgmcooling} to define an effective CGM cooling time, $\rm t_{cool,eff}$.\\
\\
The free-fall time is then determined from the gravitational acceleration at the virial radius:
\begin{equation}
    \rm t_{ff}=\sqrt{\frac{2R_{vir}}{g}}=\sqrt{\frac{2R_{vir}^3}{GM_{vir}}},
\end{equation}
where g is the gravitational acceleration, and G is the gravitational constant. The CGM cools to low temperatures and accretes onto the ISM:
\begin{equation}\label{eq:mdot_inflow}
    \dot{\rm m}_{\rm cool}=\frac{\rm m_{CGM}}{\rm t_{cool,eff}+t_{ff}},
\end{equation}
where $\rm m_{CGM}$ is the mass of the CGM gas. The two relevant timescales add in the denominator, so the slower one sets the cooling rate, $\rm \dot{m}_{cool}$.
\section{Star Formation}\label{Stars_method}
\begin{figure}
    \centering
        \includegraphics[width=\linewidth]{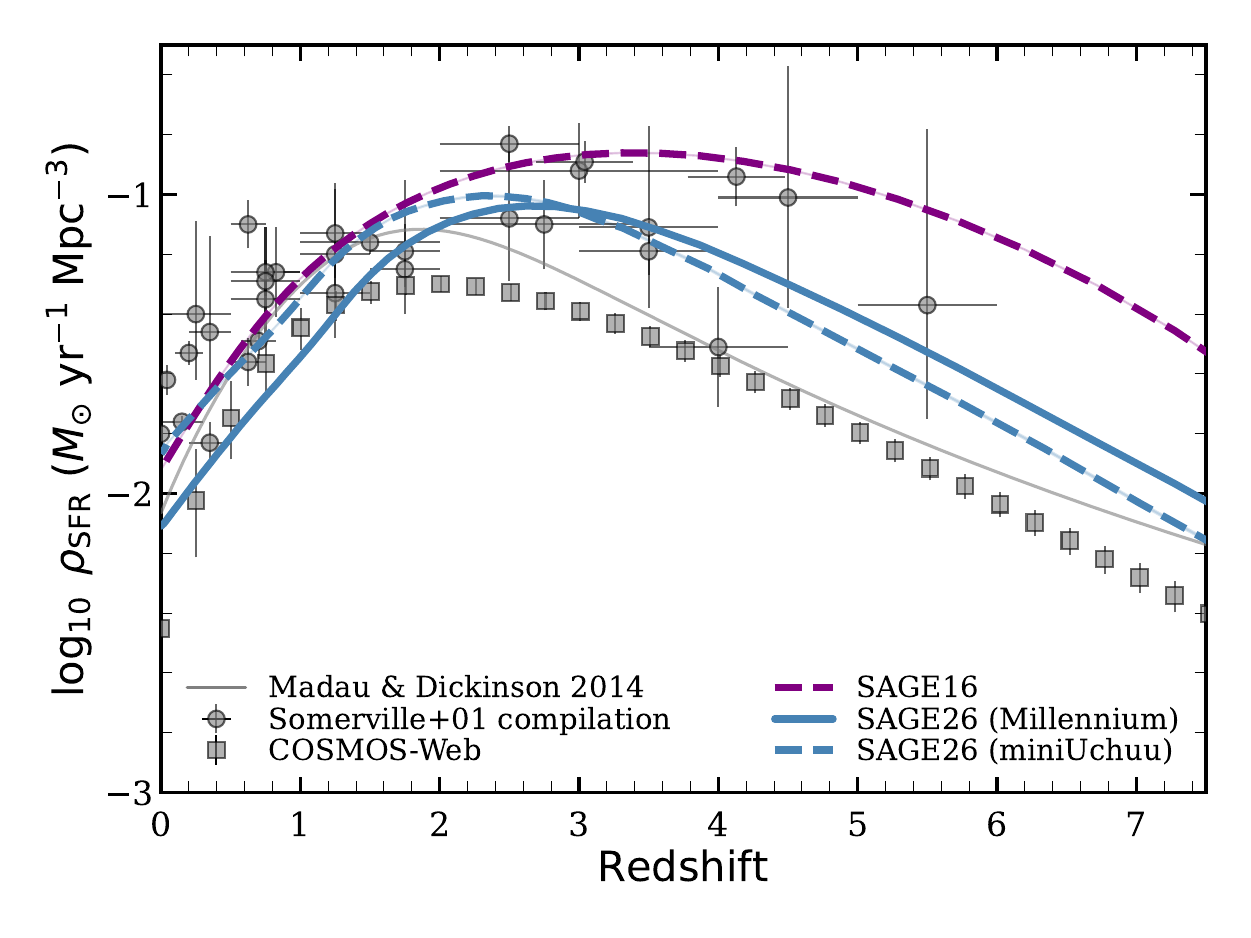}
        \caption{Cosmic star formation rate density ($\rm \rho_{SFR}$) evolution showing SAGE26's results compared with observations at high redshift. While SAGE16 (purple, dashed line) over-predicts by $\rm \sim 0.7$ dex at $\rm z=6$, SAGE26 (solid blue line for Millennium and dashed blue line for miniUchuu) reduces high-redshift star formation by a factor of $\sim3$, bringing it into agreement with the COSMOS-Web constraints of \citet{shuntov2025} (squares) while maintaining low-redshift agreement with observational scatter. We also present observations (circles) compiled by \citet{somerville2001}; see their Table A2 for a complete list of references.}
        \label{fig:sfrd}
\end{figure}
\begin{figure}
    \centering
    \includegraphics[width=\linewidth]{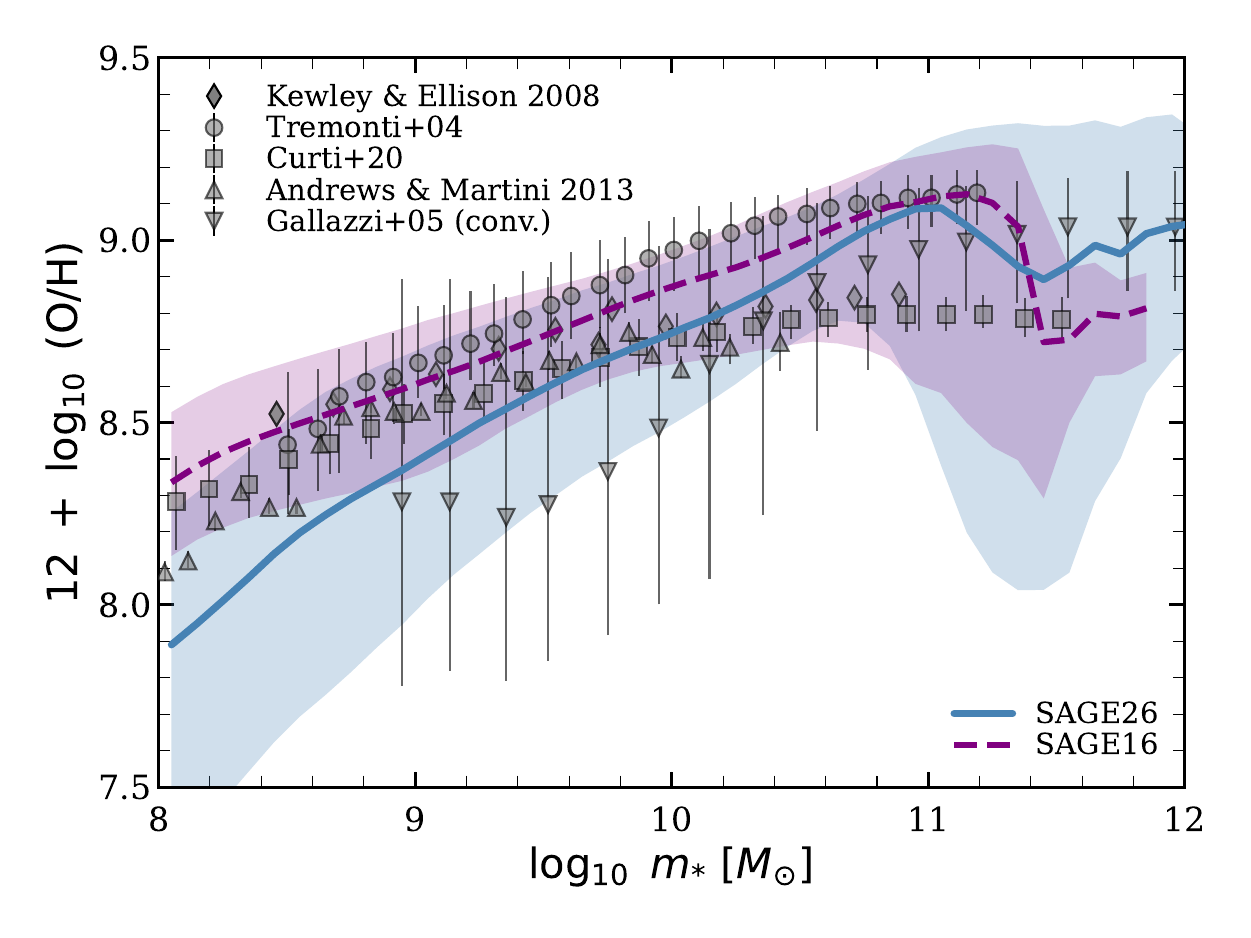}
    \caption{Mass-metallicity relation at $\rm z=0$. Solid or dashed lines represent the median relation with shaded regions enclosing the $\rm 1\sigma$ (16th-84th percentile) range, comparing SAGE26 galaxies (blue) with observations from \citet{tremonti2004,gallazzi2005,kewley2008,andrews2013,curti2020} and SAGE16 (purple). SAGE26 produces a steeper mass-metallicity relation than observed, with a median lying $0.12-0.14$ dex low at $\rm m_*\sim10^{8-9}\ M_{\odot}$ and $0.17-0.18$ dex high above $\rm m_*=10^{11}\ M_{\odot}$, but still within the scatter of observations.}
    \label{fig:metallicity}
\end{figure}

In SAGE16, stars are produced from the total cold gas mass reservoir; the addition of atomic and molecular hydrogen in SAGE26 improves the model quality, providing a more nuanced view of stellar growth within galaxies and the separation between quiescent and star-forming galaxies.\\
\\
Stars form from the rotationally supported disk in the centre of the dark matter halo, where the stellar mass of a galaxy grows as:
\begin{equation}\label{eq:6}
    \dot{\rm m}_{\ast}=\alpha_{\rm SF}\ \frac{\rm m_{H_2}}{\rm t_{dyn,disk}}.
\end{equation}
Here, $\rm m_{H_2}$ is the $\rm H_2$ mass (see \Sec{h2}), $\rm \alpha_{SF}$ is the fraction of gas that is converted to stars, per dynamical time, $\rm t_{dyn,\ disk}=r_s/V_{vir}$, with $\rm r_s$ being the disk scale radius and $\rm V_{vir}$ the halo's virial velocity. The disk radius is determined by:
\begin{equation}\label{eq:diskradius}
    \rm r_s=\frac{3}{\sqrt{2}}\lambda R_{vir},
\end{equation}
which is three times the disk scale length proposed by \cite{efstathiou1982} and \cite{mo1998}, based on the properties of the Milky Way \citep{vandenbergh2000}. The spin parameter of the dark halo, $\lambda$ \citep{bullock2001}, is taken directly from the N -body simulation.\\
\\
The updated SAGE26 model results in a galaxy population whose combined star formation rate density is more consistent with the observed Universe than its predecessor, as seen in \Fig{fig:sfrd}. Here we show the cosmic star formation rate density evolution for SAGE26 using Millennium (blue solid line) haloes, miniUchuu (blue dashed line) haloes and SAGE16 (purple dashed line) using Millennium haloes in comparison with observations. SAGE16 predicts star formation rates that are too high on average at early epochs, while SAGE26 more closely follows recent JWST observations from \citet{shuntov2025} (squares), although the uncertainties are large.\\
\\
As galaxy populations evolve, the interstellar medium becomes enriched with elements heavier than hydrogen and helium, a process known as metal enrichment. Metal enrichment is modelled by assuming that as each solar mass forms a portion of metals is produced, known as the yield via supernovae. Stars return metals to cold gas via the constant yield ($\rm Y=0.025$) and instantaneous recycling fraction $\rm R=0.43$, where R is the fraction of mass available for recycling.\\
\\
We show the metallicity of galaxies at $\rm z=0$ for SAGE26 Millennium haloes and compare with observations in \Fig{fig:metallicity}. When using the molecular hydrogen recipe for star formation, the relation is steeper than observed rather than offset. For galaxies between $\rm 10^{8-9}\ M_{\odot}$, the model median lies between 0.12 and 0.14 dex below \cite{andrews2013} and \cite{curti2020}, while galaxies above $\rm 10^{11}\ M_{\odot}$ the model lies $0.17-0.18$ dex above them. Both offsets are smaller than the model's $\rm 16^{th}\ to\ 84^{th}$ percentile spread at fixed stellar mass.

\subsection{$\rm H_2$ formation} \label{h2}

As the star formation rate in \Eq{eq:6} depends on the molecular hydrogen mass, $\rm H_2$, rather than the cold gas reservoir, we must also determine what fraction of a galaxy's cold gas has been converted to a molecular form. In massive, gas-rich galaxies, high midplane pressures compress the interstellar medium and allow efficient $\rm H_2$ formation. In these massive systems, most of the cold gas is available for star formation. In low-mass or metal-poor systems, lower pressures and reduced dust content limit the molecular fraction, naturally suppressing star formation even when atomic gas is plentiful \citep{blitz2006,krumholz2012a,krumholz2012,krumholz2013,gnedin2014}.\\
\\
To capture this, we implement the \cite{blitz2006} (BR06) pressure-based $\rm H_2$ formation prescription, which has been shown to track multi-phase gas partitioning in other semi-analytic frameworks \citep{popping2014,somerville2015}. Additionally, we also subtract a fixed 26\% of the cold gas when determining the $\rm H_2$ mass to account for dust, helium and ionised gas \citep{xie2017}. This empirical model relates the molecular-to-atomic gas ratio to the hydrostatic midplane pressure of the galactic disk:
\begin{equation}\label{eq:rmol}
    \rm R_{mol}=(P_{ext}/P_0)^{\alpha_{H_2}},
\end{equation}
where $\rm P_{ext}$ is the midplane pressure of the disk, $\rm P_0$ is a characteristic pressure scale where we use the central value of $\rm P_0=4.3\times10^4\ K\ cm^{-3}$, and $\rm \alpha_{H_2}=0.92$ is a pressure-related parameter, where $\rm P_0\ and\ \alpha_{H_2}$ follow BR06, which are calibrated from observations of nearby galaxies \citep{blitz2006}.
\begin{figure}
    \centering
    \includegraphics[width=\linewidth]{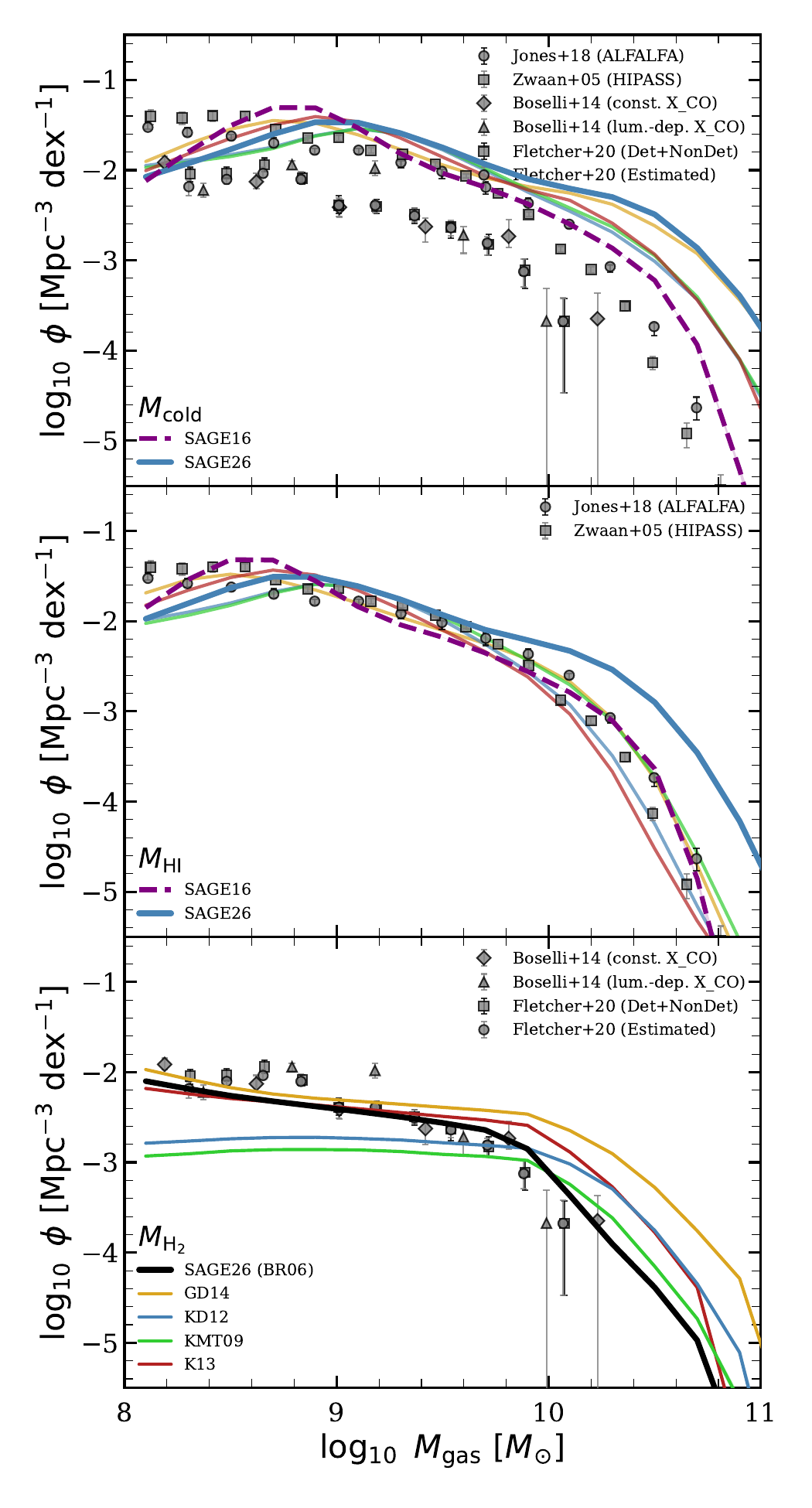}
    \caption{Cold gas (top), atomic hydrogen (middle) and molecular hydrogen mass functions at $\rm z=0$ comparing five $\rm H_2$ formation prescriptions: \citet{blitz2006} (BR06) in black, \citet{krumholz2009} (KMT09) green, \citet{krumholz2012} (KD12) blue, \citet{krumholz2013} (K13) red, \citet{gnedin2014} (GD14) yellow, SAGE16 in purple and SAGE26 in blue. We compare with \citet{zwaan2005,boselli2014a,jones2018} and \citet{fletcher2020} in the cold gas, HI and $\rm H_2$ mass functions. All models use the same galaxy population, only differing in $\rm H_2$ recipes. SAGE26 with \citet{blitz2006} $\rm H_2$ formation recipe shows good overall agreement with observations for the $\rm H_2$ mass function and is the recipe SAGE26 was calibrated with.}
    \label{fig:h2_massfunction}
\end{figure}
We calculate the midplane pressure from the cold gas and stellar surface densities within an approximation of the star-forming region, using the stellar scale height (following \citealt{blitz2006}). The molecular fraction\footnote[1]{This is calculated as a function of radius assuming exponential surface density profiles for gas and stars, and then integrated over the disk, resulting in the total molecular fraction for the galactic disk.} is then:
\begin{equation}
    f_{\rm mol}=\rm \frac{R_{mol}}{1+R_{mol}}.
\end{equation}
This pressure-based approach accounts for the observed transition from an atomic-dominated to a molecular-dominated interstellar medium as galaxies become more massive and gas-rich. Critically, it suppresses star formation in low-mass, low-surface density galaxies where pressures are insufficient for substantial $\rm H_2$ formation.\\
\\
We have tested other physical recipes for the formation of molecular hydrogen from \cite{krumholz2009,krumholz2012,krumholz2013} which all include some form of self-shielding, protecting the inner molecular hydrogen from being dissociated by far-UV photons; \cite{krumholz2012} introduces a metallicity dependence through the dust optical depth, and \cite{krumholz2013} extends this framework to the low-surface density and low-metallicity populations. We also tested the \cite{gnedin2014} recipe, which introduces dust-to-gas ratios and turbulence or clumping factors.\\
\\
In \Fig{fig:h2_massfunction} we show the cold gas, atomic and molecular hydrogen mass functions at $\rm z=0$ as a sensitivity test, comparing SAGE26 with these different prescriptions against the observations of \citet{jones2018,zwaan2005,boselli2014a,fletcher2020} and models from \cite{krumholz2009,krumholz2012,krumholz2013,gnedin2014}. All models use the same SAGE26 galaxy population but differ in the $\rm H_2$ recipe.\\
\\
Different prescriptions fail at opposite ends of the stellar mass function (not shown here). \cite{krumholz2012} and \cite{krumholz2009} reintroduce the dwarf quiescent galaxy excess and suppress the low-mass end of the $\rm H_2$ mass function. \cite{gnedin2014} and \cite{krumholz2013} preserve the star-forming and quiescent galaxies but overproduce at the massive end of the stellar mass function. BR06 is the least problematic prescription and requires no metallicity or dust tracking. Against high-mass observations, SAGE26 overpredicts above $\rm 10^{9.3}\ M_{\odot}$, carried mostly by central galaxies, and because the molecular fraction is pressure-regulated, it accumulates in the atomic phase, leaving the $\rm H_2$ mass function unaffected. At a fixed gas mass of $\rm 10^{9.5}\ M_{\odot}$, both atomic and molecular hydrogen are offset by the same amount, indicating an excess of cold gas rather than incorrect partitioning of it between phases.\\
\begin{figure}
    \centering
    \includegraphics[width=\linewidth]{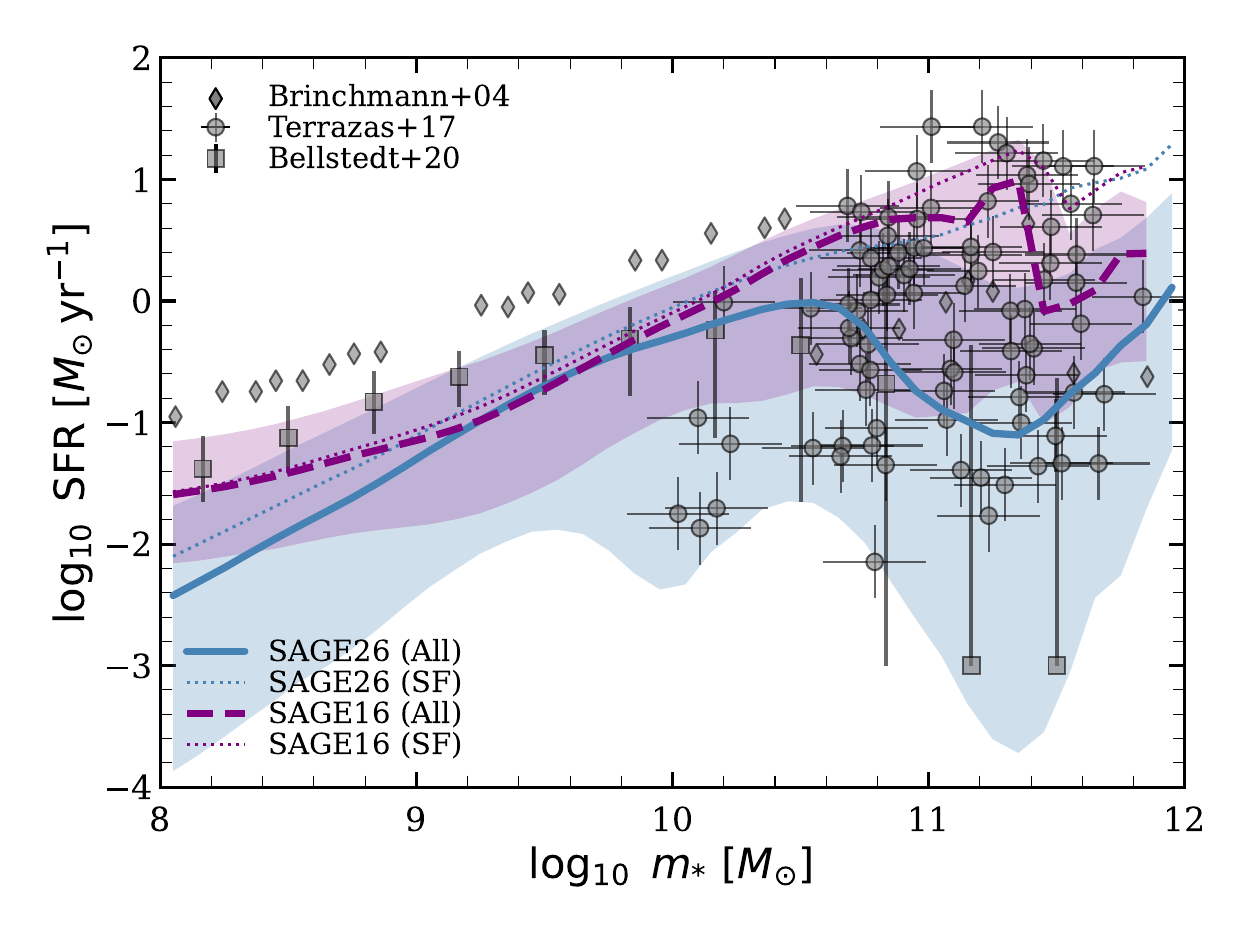}
    \caption{Star formation rate as a function of stellar mass at $\rm z=0$ for both star-forming and quiescent galaxies. Solid or dashed lines show the median star formation rate with shaded regions enclosing the $\rm 1\sigma$ (16th-84th percentile) range. SAGE26 (blue) follows the same trends as SAGE16 (purple) for low-mass galaxies. We also include the star-forming population for SAGE26 and SAGE16 with dotted lines. For high-mass galaxies, SAGE26 produces a larger abundance of quiescent galaxies. We also show observations from \citet{brinchmann2004, bellstedt2020} and \citet{terrazas2017}.}
    \label{fig:sSFR}
\end{figure}
\\
The relationship between star formation rate and stellar mass at $\rm z=0$, shown in \Fig{fig:sSFR}, shows that SAGE26 reproduces a mass dependence similar to SAGE16. However, SAGE26 now produces a substantial low star formation rate population specifically for galaxies $\rm >10^{10.5}\ M_{\odot}$, which was lacking in SAGE16. The updated model also produces many more low-mass galaxies with high star formation rates, reducing the overproduction of dwarf-quiescent seen galaxies in SAGE16.
\section{Supernova Feedback}\label{SN_method}
\begin{figure*}
    \centering
    \includegraphics[width=\linewidth]{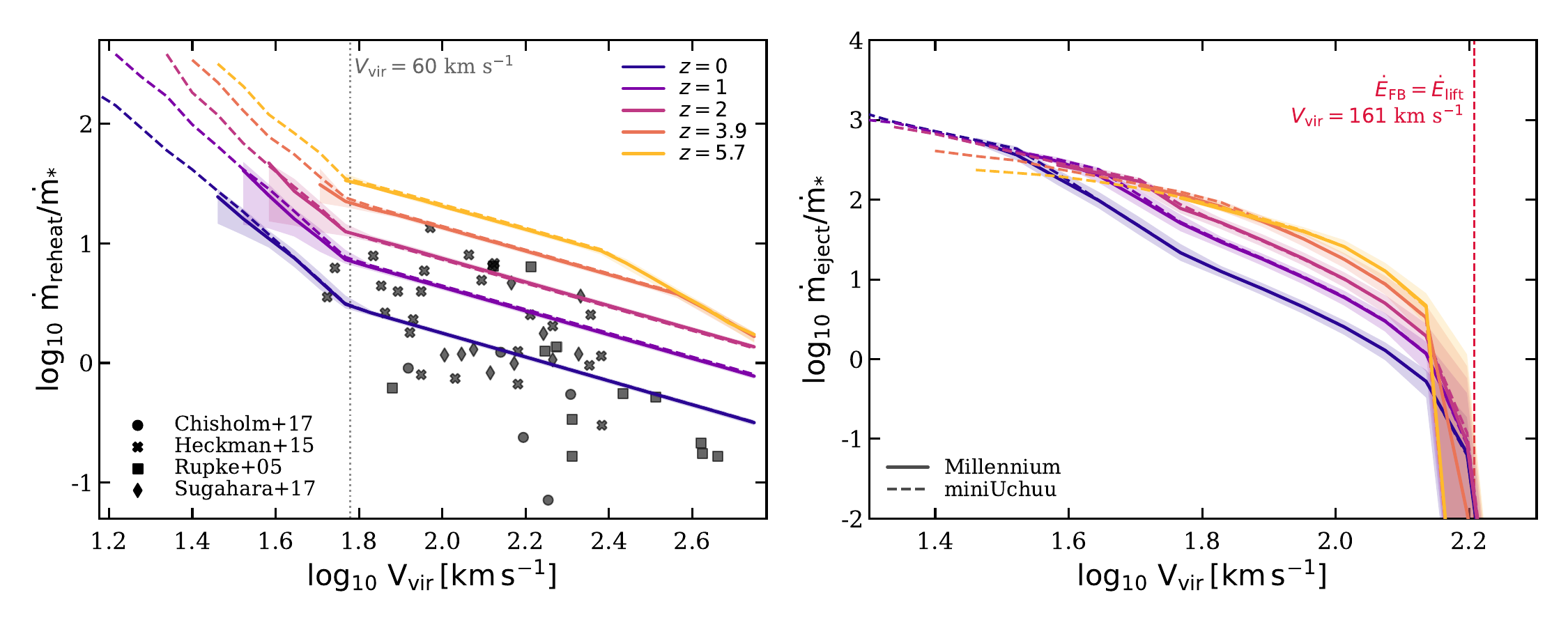}
    \caption{\textit{Left}: mass loading factor $\rm \eta_{reheat}$ at $\rm z=0,\ 1,\ 2,\ 3.9\ and\ 5.7$, with SAGE26 medians with $\rm 1\sigma$ shading for Millennium (solid lines) and miniUchuu (dashed lines) haloes as well as observations from \citet{rupke2005,heckman2015,chisholm2017} and \citet{sugahara2017}. Observations from \citet{rupke2005,heckman2015} and \citet{chisholm2017} are from $\rm z<0.5$ and \citet{sugahara2017} observations span $\rm z=0-2$. The dotted vertical line ($\rm V_{vir}=60\ km\ s^{-1}$) marks the break between the two FIRE slopes. \textit{Right}: shows the same lines and redshifts as the mass-loading panel but for ejection ($\rm \dot{m}_{eject}/\dot{m}_*$) with no ejection above $\rm 161\ km\ s^{-1}$ at every redshift.}
    \label{fig:massloading}
\end{figure*}
\begin{figure}
    \centering
    \includegraphics[width=\linewidth]{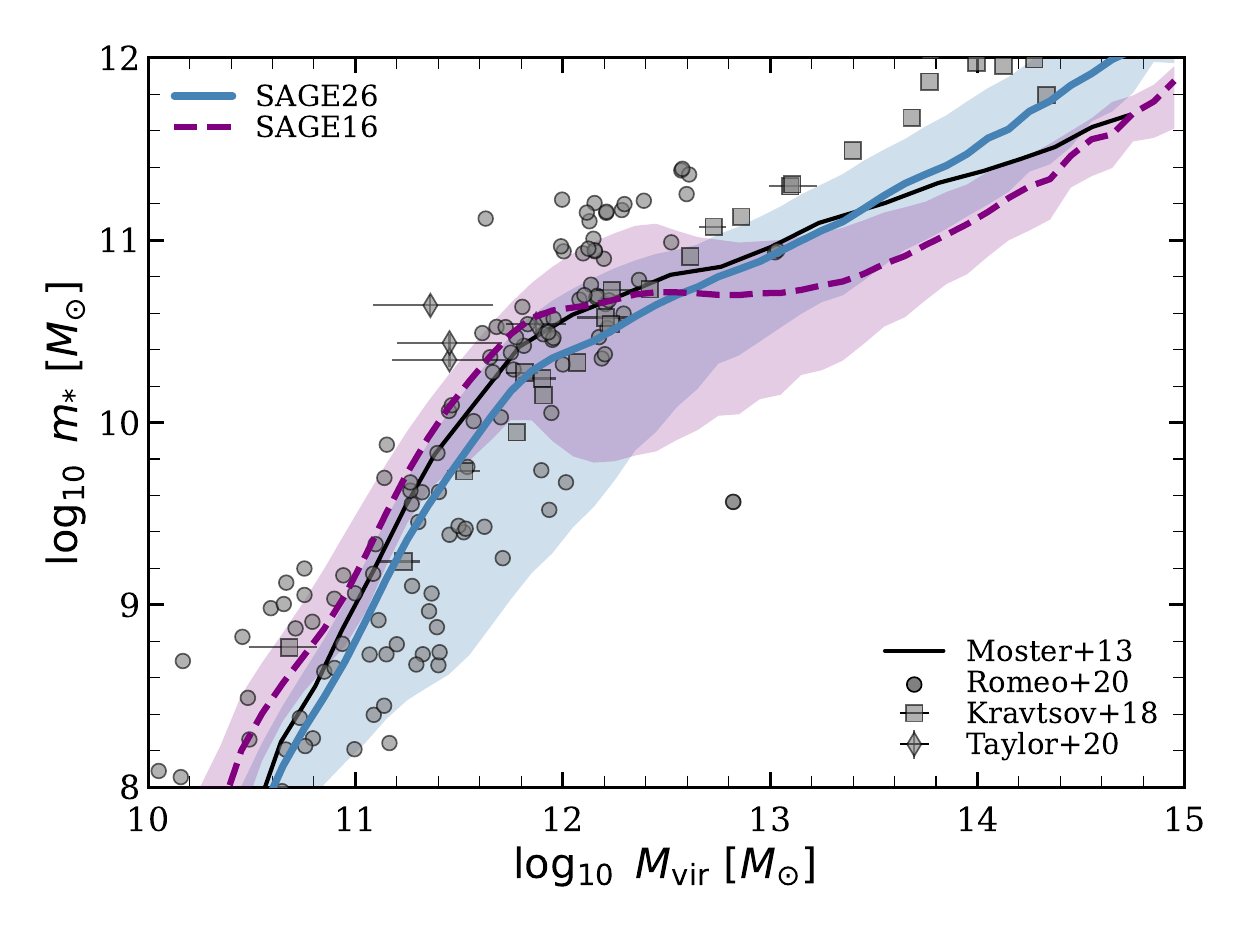}
    \caption{Stellar mass-halo mass relation at $\rm z=0$. Solid and dashed lines show the median relation, with shaded regions enclosing the $\rm 1\sigma$ (16th-84th percentile) range for SAGE26 (blue) and SAGE16 (purple), compared with observational constraints. SAGE26 shows improvement over SAGE16, reducing stellar mass over-prediction by 0.3-0.4 dex for $\rm M_{vir} \sim 10^{10.5}\ M_{\odot}$ and maintaining agreement across four orders of magnitude in halo mass, from dwarf galaxies to massive clusters. For comparison, we show observations from \citet{moster2013,kravtsov2018,taylor2020} and \citet{romeo2020}.}
    \label{fig:smhm}
\end{figure}

While molecular hydrogen prescriptions determine when and where gas can form stars within galaxies, stellar feedback controls what happens thereafter, regulating future star formation and, consequently, future galactic growth. This occurs through reheating and ejection processes from supernovae, which depend critically on the galaxy and the wider halo's gravitational potential and mass.\\
\\
High-resolution hydrodynamical simulations from the FIRE project have provided crucial insights into the mechanisms that may drive this observed mass dependence \citep{muratov2015,hirschmann2016,hopkins2018,hopkins2020,hopkins2022}. These simulations track supernova outflows with their cumulative effects over cosmic time, revealing that feedback efficiency depends critically on the ratio of supernova energy to the gravitational binding energy.\\
\\
The observational evidence points to a clear physical picture: supernova explosions can efficiently drive gas out of shallow gravitational potential wells but struggle to overcome the deeper wells of massive haloes \citep{rupke2005,heckman2015,somerville2015,chisholm2017}.\\
\\
To capture the observed mass dependence suggested by both observations and high-resolution simulations, we replace the fixed mass-loading factor of SAGE16 with one that additionally depends on the potential depth and redshift:
\begin{equation}\label{eq:etareheat}
    \rm \eta_{reheat}=\epsilon_{disk}\ (1+z)^{1.25}\ \bigg(\frac{V_{vir}}{60 \ km\ s^{-1}}\bigg)^{\alpha}.
\end{equation}
\\
We adopt the functional form of the mass-loading from \cite{muratov2015}, with the velocity break at $\rm 60\ km\ s^{-1}$ reflecting the transition from energy-driven (low-mass) to momentum-driven (high-mass) winds.\\
\\
The $\rm (1+z)^{1.25}$ redshift evolution, taken directly from FIRE simulations and \cite{hirschmann2016}, captures the evolution of feedback efficiency of higher gas densities at early epochs. The normalisation $\rm \epsilon_{disk}=2.9$ is the reheating efficiency parameter, and z is the redshift.\\
\\
$\rm \alpha$ is a velocity-dependent power-law index, for low-mass galaxies, a steep scaling reflects when supernovae are powerful relative to the host galaxy. For high-mass systems, a shallow branch reflects when supernovae are not as dominant and follow a $\rm \eta_{reheat}\propto V^{-1}$ behaviour expected of momentum-driven winds:
\begin{equation}
    \rm \alpha = \begin{cases}
    -3.2 & \text{if } \rm V_{vir} < 60 \text{ km s}^{-1} \\
    -1.0 & \text{if } \rm V_{vir} > 60 \text{ km s}^{-1}.
    \end{cases}
\end{equation}
The reheated gas due to supernova activity is then calculated with:
\begin{equation}
    \dot{\rm m}_{\rm reheat}=\eta_{\rm reheat}\ \dot{\rm m}_{*}, 
\end{equation}
where $\dot{\rm m}_{*}$ is the star formation rate for the current timestep. By varying the mass-loading with halo virial velocity, we ensure that dwarf galaxies experience high reheating rates, while massive galaxies retain their gas supply. We require that the reheating and ejection together do not use more energy than supernova release and that $\rm \dot{m}_{reheat}$ doesn't reheat more than the available gas. Reheated gas is then removed from the interstellar medium and funnelled into the CGM or the hot intragroup or intracluster medium.\\
\\
However, reheating is only the first stage of supernova behaviour. Following SAGE16 with additions from \cite{guo2011} and \cite{hirschmann2016}, we implement energy-based ejection where supernova energy must first lift reheated gas to the virial radius before any additional gas can be ejected. The total energy from supernova feedback is as follows:
\begin{equation}\label{eq:EFB}
     \dot{\textrm  E}_{\rm FB}=\epsilon_{\rm halo}f\frac{1}{2}\ \dot{\textrm m}_{*}\rm V_{SN}^2,
\end{equation}
where $\rm \epsilon_{halo}=0.3$ is the ejection efficiency, $f=\rm (1+z)^{1.25}\ (V_{vir}/60 \ km\ s^{-1})^{\alpha}$ and $\rm V_{SN}^2=\eta_{SN}\,E_{SN}$ is the mean supernova energy per unit mass of stars formed. With $\rm \eta_{SN}=5\times10^{-3}\ M_{\odot}^{-1}$ and $\rm E_{SN}=10^{51}\ erg$ this gives $\rm V_{SN}=501\ km\ s^{-1}$.\\
\\
Gas can only be ejected from the hot halo or circumgalactic medium if supernova energy exceeds the energy required to lift the reheated gas:
\begin{equation}\label{eq:Elift}
    \dot{\textrm  E}_{\rm lift}=\frac{1}{2}\ \dot{\textrm m}_{\rm reheat}\ V_{\rm vir}^2.
\end{equation}
We then calculate the ejected mass as:
\begin{equation}
    \dot{\textrm m}_{\rm eject}=\frac{\dot{\textrm  E}_{\rm FB}-\dot{\textrm E}_{\rm lift}}{\frac{1}{2}\rm V^2_{vir}}.
\end{equation}
The resulting mass-loading and ejection behaviour is shown in \Fig{fig:massloading}. On the left is the mass-loading as a function of $\rm V_{vir}$ where we show $\rm \eta_{reheat}$ at $\rm z=0,\ 1,\ 2,\ 3.9\ and\ 5.7$ with $\rm 1\sigma$ shading from SAGE26 using Millennium (solid lines) and miniUchuu (dashed lines) haloes, as well as observations (grey points) from \cite{rupke2005,heckman2015,chisholm2017} and \cite{sugahara2017}. The dotted line indicates the $\rm V_{vir}=60\ km\ s^{-1}$ break in \Eq{eq:etareheat}.\\
\\
A halo with $\rm V_{vir}=50\ km\ s^{-1}$ reheats $\rm 5.6\ M_{\odot}$ of cold gas per solar mass of stars formed at $\rm z=0$. Above the break, a $\rm V_{vir}=120\ km\ s^{-1}$ halo reheats $\rm 1.5\ M_{\odot}$, and at high velocities we can see the curves converge as the available supernova energy sets the mass-loading and the limit depends on $\rm V_{vir}$.\\
\\
The right panel shows the ejected mass per solar mass of stars formed with the same redshifts and lines as the $\rm \eta_{reheat}$ panel. This confirms that ejection is confined to shallower potentials and becomes ineffective at removing gas in more massive haloes.\\
\\
The velocity- and redshift-dependent feedback prescription naturally reproduces the observed stellar-to-halo mass relation, as shown in \Fig{fig:smhm}. At low halo masses ($\rm M_{vir}\sim10^{10.5}\ M_{\odot}$) efficient feedback, $\rm \eta_{reheat}\approx1-15$, suppresses stellar masses compared to SAGE16 bringing SAGE26 in agreement with observations. At higher halo masses ($\rm M_{vir}>10^{13}\ M_{\odot}$), reduced stellar feedback efficiency allows continued stellar mass growth until active galactic nuclei begin to dominate growth suppression, while SAGE16 appears to decline in stellar mass before increasing at very high halo masses. The star formation rate-weighted mass loading rises from $\rm \langle\eta_{reheat}\rangle=1.8$ at $\rm z=0$ to 19.6 at $\rm z=8.5$.
\begin{figure}
    \centering
    \includegraphics[width=\linewidth]{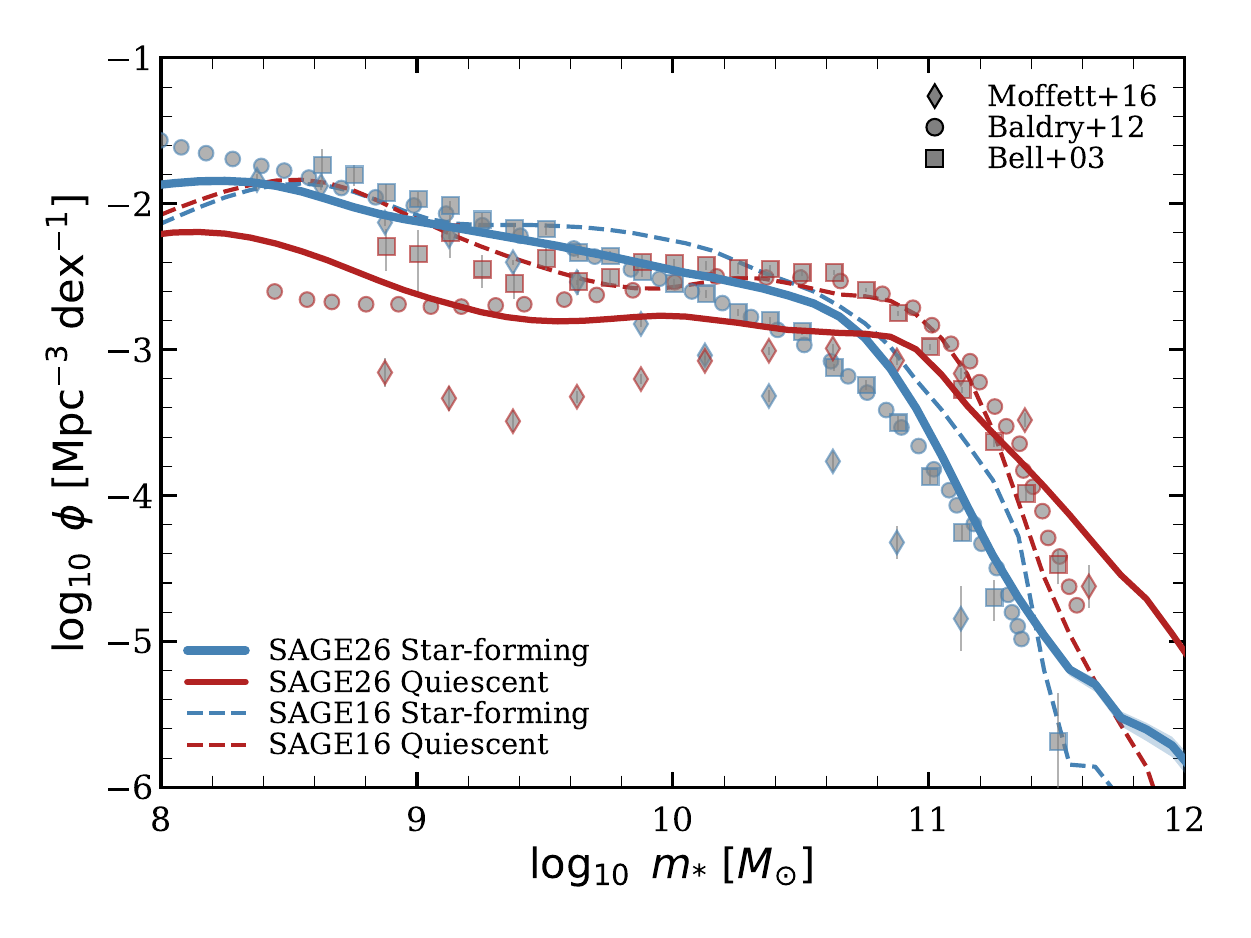}
    \caption{The stellar mass function for star-forming galaxies at $\rm z=0$ for SAGE26 (blue, solid line), SAGE16 (blue, dashed line), quiescent galaxies (quiescence defined as galaxies with specific star formation rates, $\rm log_{10}\ sSFR<-11\ yr^{-1}$) for SAGE26 (red, solid line), SAGE16 (red, dashed line) and observations from \citet{bell2003,baldry2012} and \citet{moffett2016}. SAGE26 shows an improvement at the low-mass end of the stellar mass function for quiescent galaxies, where it now more closely follows observations.}
    \label{fig:sf_q_smf}
\end{figure}\\
\\
Our model imposes a redshift-independent threshold, $V_{\rm SN}\sqrt{\epsilon_{\rm halo}/\epsilon_{\rm disk}}=161\,{\rm km\,s^{-1}}$, above which no gas can be ejected. Below this threshold, at $\rm z=0$, an $\rm 80\ km\ s^{-1}$ halo can eject $\rm 7\ M_{\odot}$ per solar mass formed and $\rm 71\ M_{\odot}$ at $\rm z=5.7$. Reheating operates at all masses while ejection does not. It is this, together with the reincorporation threshold (see \Sec{Reinc_method}), that sets the low-mass slope of the stellar-to-halo mass relation.\\
\\
We can see the ramifications of the inclusion of molecular hydrogen and the updated supernova feedback recipe on the galaxy population in \Fig{fig:sf_q_smf}, where we show the stellar mass function for star-forming and quiescent galaxies from SAGE26 (with quiescence defined as galaxies with specific star formation rates, $\rm log_{10}\ sSFR<-11\ yr^{-1}$). We also include SAGE16 and observations from \cite{bell2003,baldry2012} and \cite{moffett2016}. The $\rm H_2$ implementation, being based on the surface density of the stellar disk, ensures that low-mass galaxies, which SAGE16 previously predicts to be quiescent, are in fact star-forming \citep{harrold2024a}. The quiescent population's number density in low-mass galaxies is severely reduced.\\
\\
We also note a change in behaviours at the higher-mass end of the stellar mass function: SAGE26 now produces substantially more quiescent galaxies than star-forming galaxies. The quiescent fraction rises from 52.5\% at $\rm 10^{10.25}\ M_{\odot}$ to 84.5\% at $\rm 10^{11.25}\ M_{\odot}$, and then to 94.4\% above $\rm 10^{11.75}\ M_{\odot}$, reproducing the observed transition mass near $\rm 10^{10.5}\ M_{\odot}$ and the high-mass plateau \citep{bell2003,moffett2016}. We discuss the consequences of removing either the updated supernova feedback recipe or the molecular hydrogen in \Sec{interactions} and show the results of removing the recipes in \Fig{fig:ablation}. Overall, SAGE26 shows qualitative improvement over the SAGE16 star-forming and quiescent population predictions for the local Universe.

\section{The gas cycle: Intermediate-to-high mass systems}
Massive galaxies, groups and clusters of galaxies are treated similarly to SAGE16, where we assume the infalling gas undergoes shock heating (when $\rm M_{vir} \geq M_{shock}$) and forms a hot halo with an isothermal density profile at the virial temperature. This becomes the hot gas reservoir, seen observationally as the intragroup or intracluster medium. Gas then cools onto the galaxy at a rate dependent on the cooling time.
\subsection{Hot Halo Cooling and Cold Streams}\label{Coldstreams_method}
As a halo grows and evolves past $\rm M_{shock}$, it transitions from the free-fall-limited inflow from the circumgalactic medium to a slower hot halo accretion mode. However, this transition does not instantly cut off the cold gas supply. Even as the shock-heated atmosphere develops, cold streams can still penetrate through and deliver gas to the galaxy. At this point, a halo can experience both hot accretion and cold streams delivering gas to the galactic disk \citep{kere2005,dekel2006,dekel2009}.\\
\\
The radiative cooling  of hot halo gas proceeds via the standard SAGE16 prescription:
\begin{equation}\label{eq:coolrate}
    \dot{\rm m}_{\rm cool}=\rm \frac{1}{2}\left( \frac{r_{cool}}{R_{vir}}\right)\left( \frac{m'_{hot}}{t_{cool}}\right).
\end{equation}
Here $\rm r_{cool}$ is the cooling radius, defined as the radius at which the gas cooling time is equal to the dynamical time of the system and $\rm t_{cool}$ is defined as the cooling time for a parcel of hot gas at $\rm r_{cool}$, the same as \Eq{eq:cgmcooling} but using $\rm \rho_{hot}$ for the hot gas density, which is assumed to be isothermal. Cold streams allow a portion of the accreting gas to pierce through the shock-heated atmosphere, so not all of the halo's gas is available for radiative cooling. We implement a phenomenological model inspired by \cite{dekel2006} to determine the amount of mass in the cold streams and when the streams dominate through time. The effective mass, $\rm m'_{hot}$, is the total hot gas reservoir reduced by the fraction of gas penetrating the halo through cold streams, which is expressed as:
\begin{equation}
    \rm m'_{\rm hot} \equiv (1 - \mathit{f_{\rm stream}})\, m_{hot}.
\end{equation}
$f_{\rm stream}$ is the fraction of the reservoir carried by the cold streams, and $\rm m_{hot}$ is the hot gas reservoir. Cold streams can survive when they cool faster than they are compressed, which \cite{dekel2006} express in their equation 39, and we also follow:
\begin{equation}
    \mathcal{R} \rm \equiv \left( \frac{t_{\mathrm{cool}}}{t_{\mathrm{comp}}} \right)_{\mathrm{stream}} = \left( \frac{\mathit{f}\, \mathrm{M_*}(z)}{\mathrm{M_{vir}}} \right)^{2/3} \left( \frac{\mathrm{M_{vir}}}{\mathrm{M_{shock}}} \right)^{4/3},
\end{equation}
with $\rm M_*(z)$ the Press-Schechter clustering mass and $f\sim3$. Streams can penetrate the hot halo when $\mathcal{R}<1$, and rather than applying a binary cut-off we smooth $f_{\rm stream}$:
\begin{equation}
    \rm \mathit{f}_{stream}(M_{vir},z)= \mathit{S}\left( \frac{-log\mathcal{R}}{\Delta log\mathcal{R}} \right),
\end{equation}
where $\rm \mathit{S}(x) = (1 + 10^{-x})^{-1}$ and $\rm \Delta log \mathcal{R} = 0.15$ dex, which allows both cold streams and radiative cooling to coexist. Setting $\mathcal{R}=1$ recovers a maximum mass for cold streams to occur (following \citealt{dekel2006} equation 40):
\begin{equation}
    \mathrm{M_{stream}}(z) \sim \frac{\mathrm{M^2_{shock}}}{f\, \mathrm{M_*}(z)},\ \rm \mathit{f}M_*<M_{shock}
\end{equation}
The critical redshift separating where cold streams can dominate is defined by $\rm \mathit{f}M_*(z_{crit})=M_{shock}$. Cold streams are confined to below $\rm M_{shock}$ at $\rm z<z_{crit}$ and can reach increasingly massive haloes above it with no explicit redshift cut.\\
\\
The hot gas reservoir is partitioned into gas that cools radiatively using the traditional SAGE16 prescription and cold streams which cool on the faster dynamical timescale:
\begin{equation}
    \dot{\rm m}_{\rm stream} = \frac{\mathit{f_{\rm stream}}\,\rm  m_{hot}}{\rm t_{\rm dyn}},
\end{equation}
where $\rm t_{dyn}=R_{vir}/V_{vir}$. The total cooling rate onto the galaxy in the hot regime is thus:
\begin{equation}
    \dot{\rm m}_{\rm total} = \dot{\rm m}_{\rm cool} + \dot{\rm m}_{\rm stream}.
\end{equation}
When $f_{\rm stream}=0$, we recover the SAGE16 hot-accretion-mode cooling prescription; below $\rm M_{shock}$ there is no hot atmosphere for cold streams to penetrate, and the inflow model of \Sec{CGM_method} applies. Cold streams deliver all of the gas at $\rm z=4$; $\rm M_{stream}$ then falls with $\rm M_*(z)$ and radiative cooling takes over, competing only in the most massive haloes by $\rm z=2$. By $\rm z=1$ the crossover appears at $\rm log_{10}(V_{vir}/km\ s^{-1})\approx 2.1$ where only the smallest hot haloes have cold streams. The residual streams can appear at $\rm z=0\ and\ 1$, these are haloes lying below $\rm M_{stream}$, which the stochastic regime assignment scatters across $\rm M_{shock}$.
\section{Black holes and Active Galactic Nuclei}\label{AGN_method}
The cold gas delivered through both radiative and cold stream accretion can fuel not only star formation but also the growth of central supermassive black holes. The resulting AGN activity subsequently regulates the growth of massive systems.\\
\\
AGN are incorporated into SAGE26 to play a further critical role in the life cycle of such galaxies. Following established paradigms \citep{croton2006}, AGN feedback transports energy, gas and metals within the halo and can expel mass completely, reducing the amount of cold gas available for star formation. SAGE26 follows the same method as SAGE16 and so has two modes when categorising AGN: (1) the radio mode, a low-energy long-term heating of halo gas, and (2) the quasar mode, a high-energy short-term heating and ejection of disk, halo gas and metals.
\subsection{Radio Mode AGN}
In radio mode, SAGE26 uses the Bondi-Hoyle formula to model the accretion of hot gas onto the central black hole. This describes the accretion rate as a function of the local temperature and the black hole mass, which is seeded during merger events:
\begin{equation}\label{eq:12}
    \dot{\rm m}_{\rm BH,R}=\kappa_{\rm R}\, \frac{15}{16}\, \pi\, \rm G\, \bar{\mu}m_{p}\, \frac{\rm kT_{vir}}{\Lambda}\, m_{BH},
\end{equation}
where G is the Gravitational Constant and $\kappa_{R}$ is a radio mode efficiency parameter.\\
\\
The accretion rate given by \Eq{eq:12} allows for the estimation of the black hole luminosity when in the radio mode, $\mathrm L_{\rm BH,R}=\eta\dot{\rm m}_{\rm BH,\ R}\rm c^2$, where $\rm \eta=0.1$ is the standard efficiency with which inertial mass is liberated when approaching the event horizon, and c is the speed of light. This luminosity acts as a source of heating, offsetting the energy losses from the cooling gas. If and when enough energy from the central AGN is injected into the cooling flow, the cooling can be completely turned off.\\
\\
If we define the specific energy of gas (per unit mass) in the hot halo as $\rm \frac{1}{2}V_{vir}^2$ and $\rm L_{BH,\ R}$ as the energy added, then the heating rate from radio mode feedback can be determined with:
\begin{equation}\label{eq:heatrate}
    \dot{\rm m}_{\rm heat}=\rm \frac{L_{BH,R}}{\frac{1}{2}V_{vir}^2}.
\end{equation}
Radio mode uses the following equation for the effective cooling rate, modified by the presence of the heating radius:
\begin{equation}\label{eq:13}
   \dot{\rm m}_{\rm cool}^{'}=\Biggl( 1-\frac{\rm r_{heat}}{\rm r_{cool}}\Biggl)\ \dot{\rm m}_{\rm cool}.
\end{equation}
The heating radius, $\rm r_{heat}$ is determined by finding the radius at which the energy deposited into the gas due to the radio mode equals the cooling losses interior to that radius:
\begin{equation}
    \mathrm{r}_{\rm heat}=\frac{\dot{\rm E}_{\rm heat}}{\dot{\rm E}_{\rm cool}}\rm  r_{cool},
\end{equation}where
\begin{equation}
    \dot{\rm E}_{\rm heat}=\frac{1}{2}\dot{\rm m}_{\rm heat} V^2_{\rm vir}\ \mathrm{and}\ \dot{\rm E}_{\rm cool}=\frac{1}{2}\dot{\rm m}_{\rm cool}\rm V^2_{vir}.
\end{equation}
Here $\dot{\rm m}_{\rm cool}$ is derived from \Eq{eq:coolrate} and $\dot{\rm m}_{\rm heat}$ is calculated with \Eq{eq:heatrate}.

\subsection{Quasar Mode AGN}
\begin{figure}
    \centering
    \includegraphics[width=\linewidth]{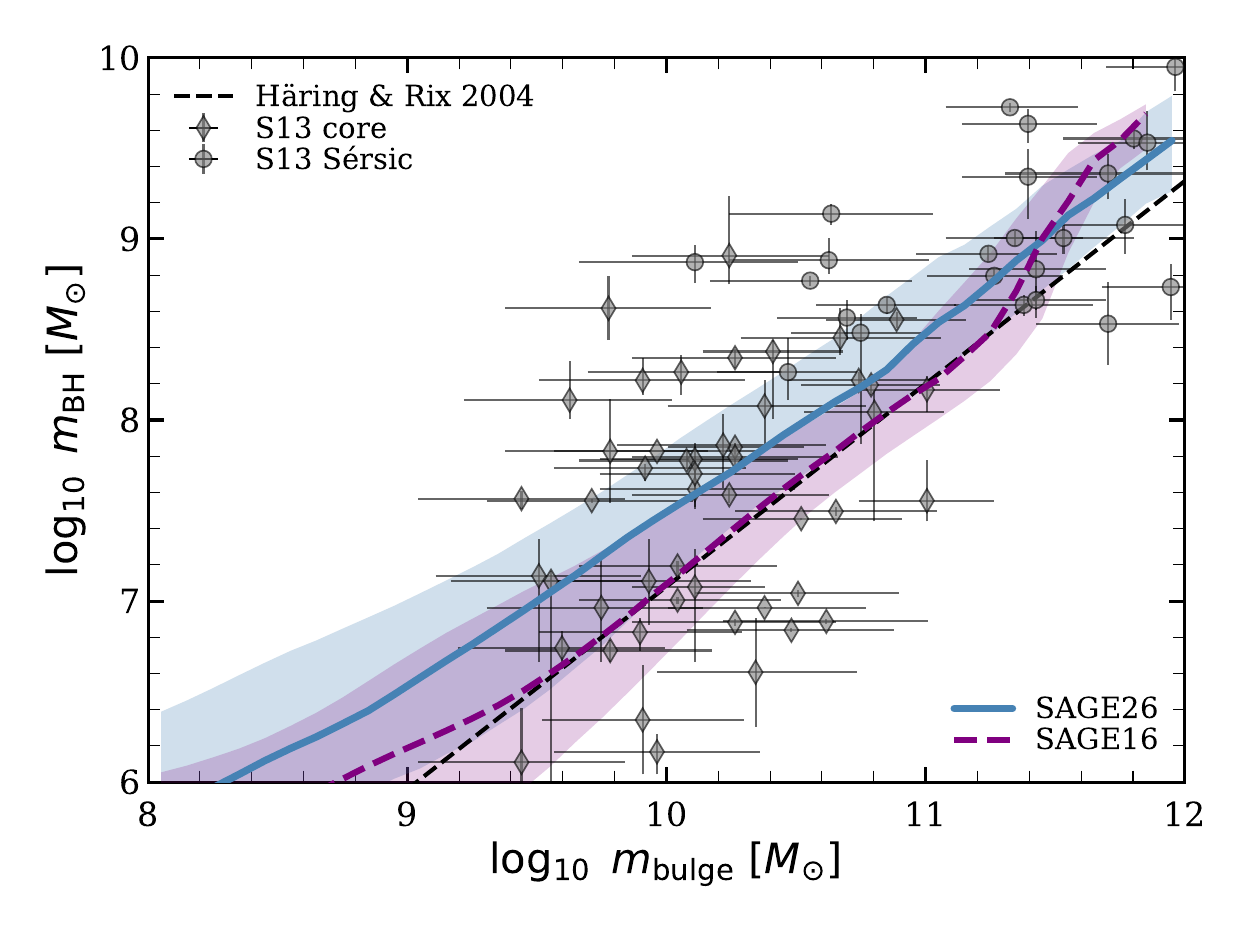}
    \caption{Bulge mass to black hole relation for a sample of galaxies at $\rm z=0$. Solid and dashed lines show the median relation with shaded regions enclosing the $\rm 1\sigma$ (16th-84th
    percentile) range for SAGE26 (blue) galaxies in comparison to Sérsic galaxies and core-Sérsic galaxies from \citet{scott2013} and the more traditional line from \citet{haering2004}.}
    \label{fig:BHBR}
\end{figure}

The second mode of black hole feedback is the quasar mode. Quasars are triggered by galaxy-galaxy mergers or by instability within the galactic disk (both described in \Sec{mergers} and \Sec{instabilities}, respectively), which funnel gas into the galactic centre very quickly. This results in high black hole accretion rates and rapid growth, making the quasar mode the dominant mass-gaining mode for supermassive black holes (see figure 3 of \citealt{croton2006}). In this model, when a merger or instability occurs, the rate at which the black hole mass grows is described by \cite{kauffmann2000}:
\begin{equation}\label{eq:15}
   \rm \Delta m_{BH,Q}=\frac{f_{BH}^{'}m_{cold}}{1+(280\, km\, s^{-1}/V_{vir})^{2}}.
\end{equation}
Here $\rm f_{BH}^{'}=f_{bh}(m_{sat}/m_{central})$ is the fraction of cold gas accreted by the black hole, determined by the ratio of the mass of the merging satellite and central galaxy. $\rm f_{bh}$ is the parameter controlling the rate of black hole growth during quasar mode. For the case of instability-driven accretion, we assume that $f'_{\rm BH}=f_{\rm BH}$ and substitute the mass of unstable gas for the cold gas mass, $\rm m_{cold}$.\\
\\
When a merger or disk instability event occurs, and the black hole has had rapid accretion, we assume a quasar wind follows with luminosity $\mathrm L_{\rm BH,Q}=\eta\dot{\rm m}_{\rm BH,Q}\rm c^2$, where $\rm \eta=0.1$ as before. This is used to calculate the total energy contained in the quasar wind:
\begin{equation}\label{eq:16}
   \rm E_{BH,Q}=\kappa _{Q}\, \frac{1}{2}\, \eta\, \Delta m_{BH,Q}\, c^{2},
\end{equation}
where $\rm \kappa_{Q}$ is a parameter of the quasar's wind efficiency. The total thermal energies for the cold disk and hot halo gas are calculated as follows:
\begin{equation}\label{eq:17}
    \rm E_{cold}=\frac{1}{2}\, m_{cold}\, V_{vir}^{2}\, and\, E_{hot}=\frac{1}{2}\, m_{hot}\, V_{vir}^{2}.
\end{equation}
If the total wind energy of \Eq{eq:16} exceeds that of the cold disk gas in \Eq{eq:17}, the gas and metals are expelled from the galaxy. If the wind energy exceeds both the cold disk and hot halo energies in \Eq{eq:17}, then all gas is ejected from the galaxy and halo along with its metals.\\
\\
In \Fig{fig:BHBR} we show that our treatment of black hole evolution is in general agreement with the local black hole–bulge mass relation and is relatively unchanged from SAGE16. We compare with the observed sample published in \cite{scott2013}, which treats Sérsic and core-Sérsic galaxies (the latter typically with more massive bulges) separately \citep{graham2014}. The observations are biased towards large bulge masses, whereas our model is complete down to much lower bulge-mass systems. The region of overlap spans more than 1 dex in width and shows clear agreement.
\section{Reincorporation}\label{Reinc_method}
SAGE26 maintains the reincorporation framework established in SAGE16, recognising that ejected gas (primarily from supernovae) may not remain ejected permanently in a dynamically evolving universe. Ejected gas can reincorporate back into the hot gas reservoir or the circumgalactic medium following a velocity-dependent prescription.\\
\\
The reincorporation rate for ejected gas follows:
\begin{equation}
    \dot{\rm m}_{\rm reinc}=\bigg(\frac{\rm V_{vir}}{\rm V_{crit}}-1\bigg)\ \frac{\rm m_{ejected}}{\rm t_{dyn}},
\end{equation}
where $\rm t_{dyn}=R_{vir}/V_{vir}$, and assuming $\rm V_{vir}>V_{crit}$ ($\rm \dot{m}_{reinc}=0$ otherwise). Here, $\rm V_{crit}=\kappa_{reinc}V_{esc}$, where $\rm V_{esc}=V_{SN}/\sqrt{2}$ is the critical halo virial velocity, above which the supernova wind velocity is insufficient for the gas to escape the halo, and $\rm \kappa_{reinc}$ parameterises how efficient this idealised process is.\\
\\
The velocity-dependent reincorporation ensures that $\rm V_{crit}$ sets the velocity scale below which no gas can reincorporate into a hot halo or the CGM, as expected for very low-mass systems where the gas has effectively escaped. Conversely, far above this scale, ejected gas reincorporates within a fraction of the dynamical time, making it quickly available in more massive systems for future cooling and, potentially, star formation.\\
\\
As noted in SAGE16, in practice, the ejected gas component need not be physically removed from the system—it merely marks gas that is unable to cool into the disk to form stars for some time.
\section{Mergers and Intracluster Stars}\label{mergers}
Galaxies are not static structures; they are dynamic and evolve through internal and external mechanisms. Morphological transformations alter the distribution of stars and gas, influence the star formation efficiency, and determine whether we see a disk-dominated or bulge-dominated galaxy.\\
\\
Observations indicate that merging galaxies, particularly satellite galaxies, can be shredded and disrupted before merging with the central galaxy, becoming part of the intragroup or intracluster stellar halo. In SAGE26, satellites experience hot-halo and CGM stripping while infalling, which occurs in proportion to dark matter subhalo stripping, and any remaining hot gas or CGM cools normally.\\
\\
To determine the satellite's ultimate fate, we need a way to decide whether it disrupts during its orbit or survives long enough to merge with the central galaxy.\\
\\
For the merging satellite, upon infall into the larger halo, an average merger time is calculated assuming dynamical friction \citep{binney2008}:
\begin{equation}\label{eq:19}
    \rm t_{friction}=1.17\, \frac{V_{vir}R_{vir}^{2}}{G\, m_{sat}\, ln\ \Lambda},
\end{equation}
\\
where $\rm m_{sat}$ is the total mass of the satellite galaxy plus subhalo system, and the Coulomb logarithm $\rm ln\ \Lambda$ can be approximated with $\rm ln(1\, +\, M_{vir}/m_{sat})$.\\
\\
\Eq{eq:19} describes the average time to merge, but every merging system will be different. As such, we track each satellite subhalo over time in the simulation, and the model measures its subhalo-to-baryonic mass ratio. If this ratio falls below a critical threshold, $f_{\rm friction}$, the satellite's survival time, $\rm t_{survive}$, is compared with the average time, $\rm t_{friction}$. If the subhalo has survived longer than the mean lifetime, the system is classified as more tightly bound and is merged with the central galaxy.\\
\\
If the subhalo does not survive the average time, the system is classified as more loosely bound, and the satellite's stars are instead disrupted and transferred to a reservoir of intracluster stars. As with SAGE16, SAGE26 produces its intracluster stellar halo primarily through mergers over time.\\
\\
Intracluster stars are stars that are no longer gravitationally bound to a host galaxy but still reside within the gravitational potential well of the dark matter halo. Observationally, intracluster stars exhibit very low surface brightness \citep{montes2022b,montes2022a, martinez-lombilla2022}, and the diffuse light emanating from them can extend far beyond the brightest cluster galaxy within a massive halo. Most groups and clusters of galaxies have some component of intracluster stars, which can be used to track the halo's evolutionary pathway. Future work will explore intracluster stars and their connection to the evolution of the host halo.\\
\\
For satellites classified as merging, the satellite-to-central galaxy baryonic mass ratio is checked. If the ratio is above the threshold $f_{\rm major}=0.3$, we class the merger as \enquote{major}, while below this threshold the merger is classified as \enquote{minor}.\\
\\
In a major merger, both galactic disks are destroyed, and the combined mass forms a spheroid or bulge. If the merger is classified as minor, then only the satellite stars are added to the central galaxy bulge. Additionally, any cold gas present can trigger a starburst, which we describe in \Sec{Starbursts}.
\section{Disk Instabilities}\label{instabilities}
Galaxies can be strongly affected and transformed due to disk instabilities, and SAGE26 follows SAGE16, which models this behaviour by measuring the Toomre Q stability criterion:
\begin{equation}\label{eq:18}
    \rm V_{c}\ /\ (G\, m_{disk}/r_{s})^{\frac{1}{2}}\, \geq\, 1.
\end{equation}
Here $\rm m_{disk}$ is the mass of the disk and $\rm r_{s}$ is the same disk radius as \Eq{eq:diskradius}. The disk is stable when this inequality is met. When the left-hand side is less than one, a portion of the stellar and cold-gas mass is transferred to the galaxy bulge to restore stability, thereby altering its morphology.\\
\\
This process can also trigger a starburst (which we discuss in the following section), with newly formed stars transferred to the bulge and part of the unstable cold gas accreted onto the central black hole.
\section{Starbursts}\label{Starbursts}\label{starbursts}
Starbursts are the rapid formation of new stars, generally triggered by an event. In SAGE26 (as in SAGE16), both mergers and disk instabilities act as triggers for starbursts.\\
\\
For merger-driven starbursts, the resulting burst occurs in proportion to the total sum of the interstellar medium in the two merging galaxies. The fraction of cold gas converted to stars is given by \cite{somerville2001}:
\begin{equation}
    \rm \mathit{e}_{burst}=\beta_{burst}(m_{sat}/m_{central})^{\alpha_{burst}},
\end{equation}
where $\rm \alpha_{burst}=0.7$ and $\rm \beta_{burst}=0.56$ are fixed parameters providing a good fit to numerical results for merger mass ratios ranging from 1:10 to 1:1 \citep{cox2004}. When the merger is minor, the new stars are added to the galactic bulge and the stability of the disk is checked. In major mergers, all stars are transferred to the spheroid, including newly formed ones. For disk instability-driven starbursts, $e_{\rm burst}$ is taken from the fraction of cold gas that is unstable, minus any gas that is accreted onto the central black hole. All newly formed stars from a disk instability are then added to the bulge, following SAGE16.
\section{The Bulge Size-Mass Relation}\label{bulges}
\begin{figure}
    \centering
    \includegraphics[width=\linewidth]{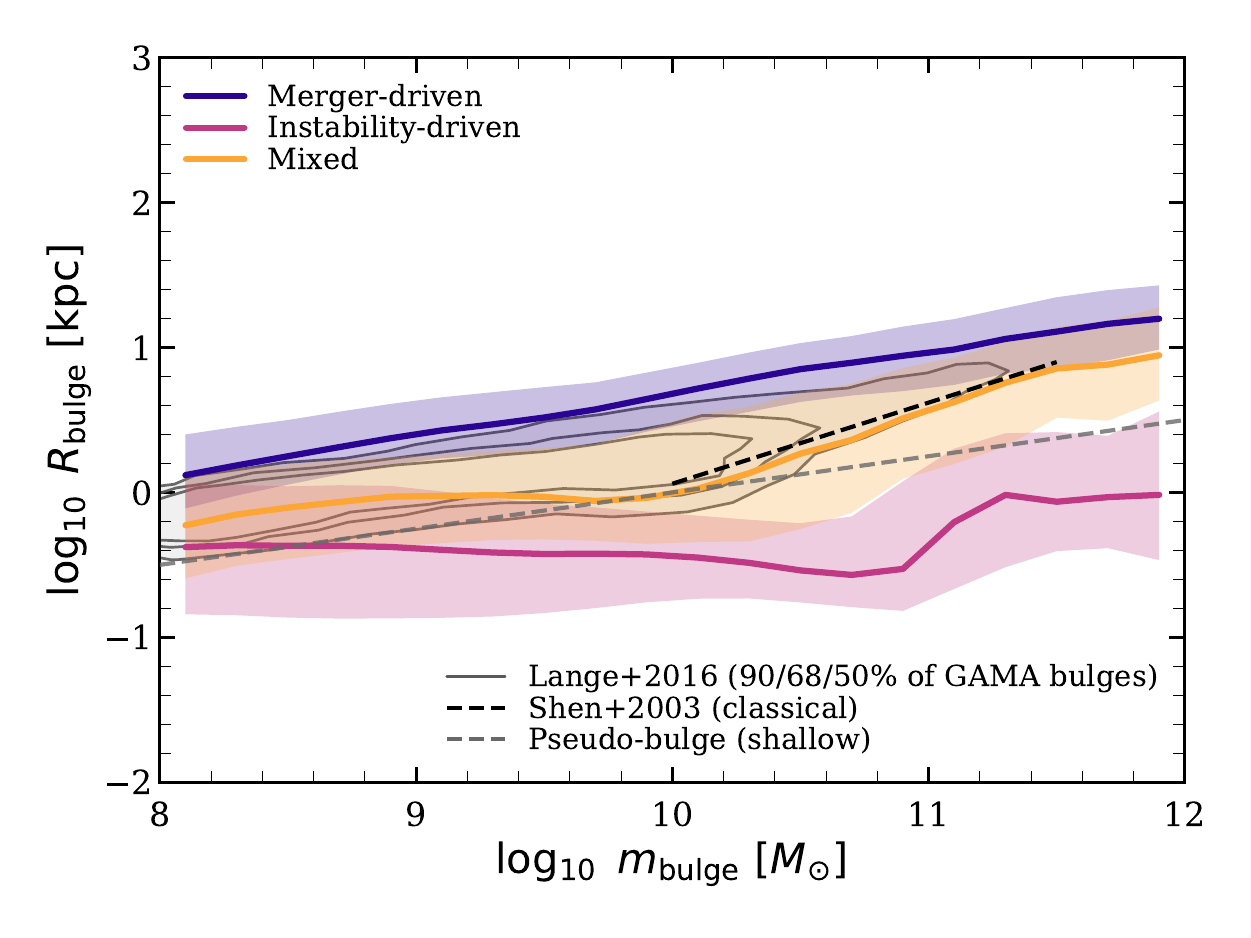}
    \caption{Bulge size-mass relation at $\rm z=0$, colour-coded by formation channel showing median lines with $\rm 1\sigma$ shading. Grey contour lines show the GAMA bulge-disk decompositions of \citet{lange2016}. The \citet{shen2003} early-type relation (black-dashed) spans the mass range it was fitted to. Merger-driven bulges follow the observed relation, lying $0.16$ dex above the \citet{lange2016} median on average; instability-driven bulges follow a shallower relation characteristic of pseudo-bulges (grey dashed), and mixed-origin bulges are rarer, with more constant radii at low masses and an upturn at higher masses.}
    \label{fig:bulgetype}
\end{figure}
Whenever a merger or disk instability event occurs, SAGE16 updates the mass of the bulge within the galaxy. SAGE26 takes this a step further by calculating and evolving its bulge radius, as first implemented by \cite{tonini2016}.\\
\\
For instability-driven bulges (also called \enquote{pseudo-bulges}; \citealt{fisher2008}), when an instability occurs and mass is transferred to the bulge component, we record the instability-driven bulge mass $\rm m_{instability}$, and from it the instability-bulge radius, $\rm r_{instability}$, evolves incrementally:
\begin{equation}
    \rm r_{instability}=\frac{r_{instability,old}\ m_{instability,old}+\delta m\times0.2r_s}{m_{instability,old}+\delta m}.
\label{eq:instabilityradiius}
\end{equation}
The effective radius of this newly transferred mass is modelled as $\rm 0.2r_s$, where $\rm r_s$ is the pre-instability disk scale radius (same as \Eq{eq:diskradius} but without the factor of 3). Because a system's disk radius scales with its specific angular momentum, the factor of 0.2 assumes that instability-driven bulges retain 20\% of the disk's specific angular momentum \citep{fisher2008,fisher2010}. This formation ensures that the pseudo-bulge's structural evolution remains tightly coupled to the properties of its host disk.\\
\\
For minor mergers, we distinguish between disk-dominated and bulge-dominated central galaxies; for disk-dominated galaxies, we determine the bulge radius using \Eq{eq:instabilityradiius}. For bulge-dominated central galaxies, we use the same approach as for major mergers.\\
\\
Since major mergers can destroy disk structure and form merger-driven bulges, following \cite{hatton2003,covington2011,porter2014} and \cite{tonini2016}, we calculate the remnant radius via energy conservation:
\begin{align}
    \rm E_{final}&=\rm E_{initial}+E_{orbital}+E_{radiative} \nonumber\\
    &=\rm G(m_1+m_2+m_{burst})^2/(2r_{final}).
\label{eq:efinal}
\end{align}
Throughout this section, binding energies are treated as positive quantities, so that the dissipative term $\rm E_{radiative}$ increases $\rm E_{final}$ and therefore reduces the remnant radius. The combined initial gravitational energy of the two progenitors is:
\begin{equation}
    \rm E_{initial}=G\left(\frac{m_1^2}{2r_1}+\frac{m_2^2}{2r_2}\right).
\end{equation}
The collisional energy is given by the orbital energy of the pair of galaxies, assuming a circular orbit:
\begin{equation}
    \rm E_{orbital}=G\ \frac{m_1m_2}{r_1+r_2}.
\end{equation}
Lastly, the radiative term takes into account the energy dissipation of the gas component due to shocks: 
\begin{equation}
    \rm E_{radiative}=C_{rad}\ E_{initial}\left(\frac{m_{gas,\ 1}+m_{gas,\ 2}}{m_1+m_2}\right).
\label{eq:radenergy}
\end{equation}
We adopt $\rm C_{rad}=2.75$ from \cite{covington2011}. For \Eq{eq:radenergy}, we use the interstellar medium cold gas for $\rm m_{gas}$, and for the total mass in the denominator, we use the baryonic mass. Solving \Eq{eq:efinal} for $\rm r_{final}$ results in the merger-driven bulge radius, $\rm r_{final}=r_{merger}$.\\
\\
We calculate the bulge half-mass radius from the mass-weighted average of both the merger and the instability radii, where they exist:
\begin{equation}
    \rm r_{bulge}=\frac{m_{merger}r_{merger}+m_{instability}r_{instability}}{m_{merger}+m_{instability}},
\end{equation}
linking galaxy structure directly to assembly history and producing mass-size relations for bulges formed through different channels.\\
\\
We show the resulting bulge size-mass relation at z=0, separated by formation channel in \Fig{fig:bulgetype}. Here, the model shows that merger-driven bulges (blue) exhibit a steeper relation that tracks the observed bulge sizes of \cite{lange2016}, lying $0.04-0.28$ dex above their median across $\rm 10^{8-10}\ M_{\odot}$. They sit 0.32 dex above the \cite{shen2003} early-type relation over the fitted range.\\
\\
The \cite{shen2003} relation is measured for early-type galaxy half-light radii against total stellar mass, and falls below the observed bulge sizes when extrapolated to lower bulge masses. In contrast, instability-driven bulges (red) follow a shallower relation similar to \enquote{pseudo-bulges}, with mixed bulges (yellow) being rarer and showing more constant bulge radii at lower masses with an upturn at higher masses. The \cite{lange2016} sample is not separated by formation channel, so the comparison is most meaningful for the merger-driven population.
\section{Feedback-Free Bursts}\label{FFBs_method}
\begin{figure}
    \centering
    \includegraphics[width=\linewidth]{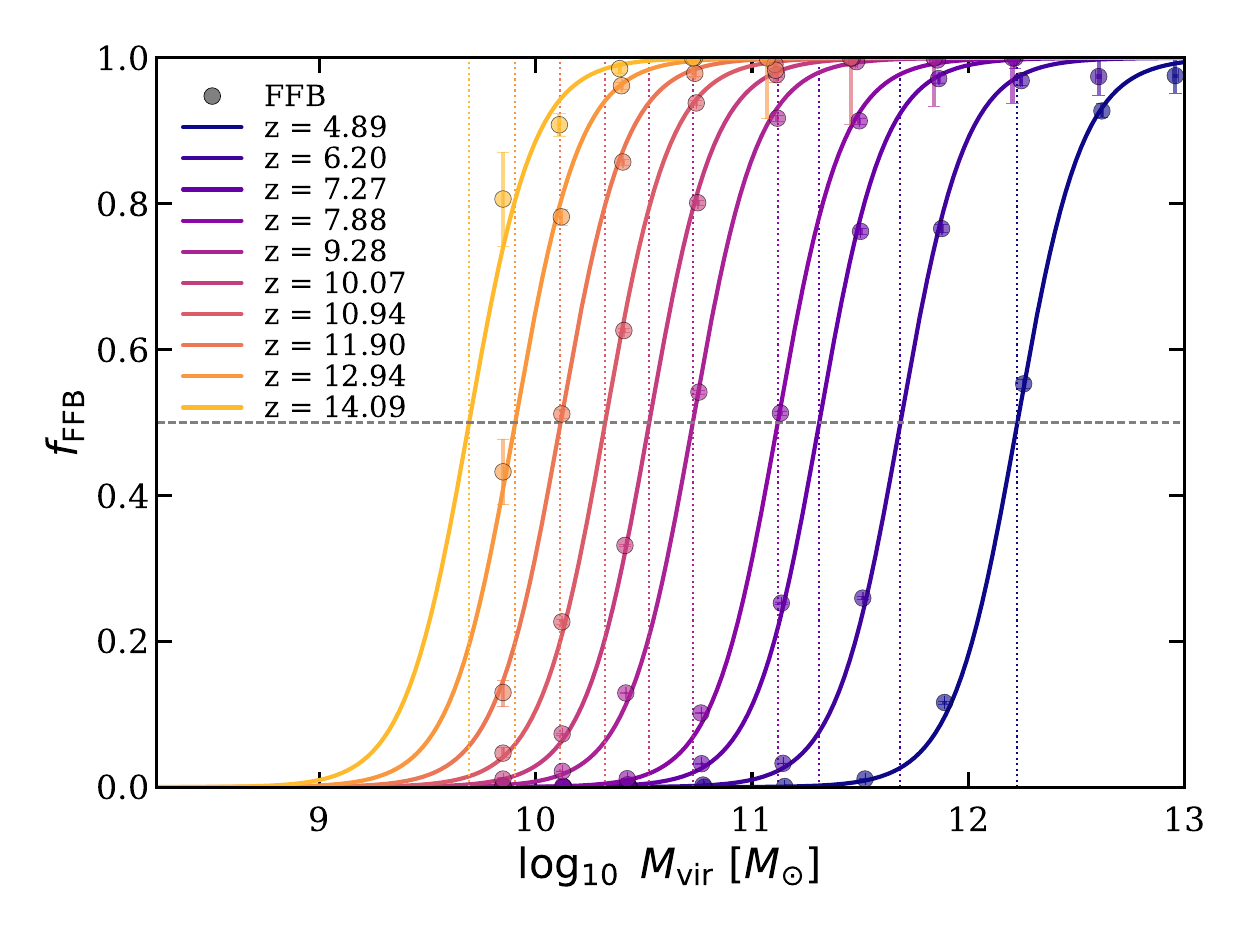}
    \caption{Fraction of galaxies in the feedback-free burst regime as a function of halo mass. Solid lines represent the \citet{li2024} model, and markers show SAGE26 haloes with bootstrap errors. The vertical dashed lines indicate the threshold mass without the sigmoidal smoothing at each redshift, where with the smoothing, $f_{\rm FFB}=0.5$. As redshift increases, the threshold mass for feedback-free burst galaxies decreases, extending to lower-mass haloes at earlier epochs of the Universe. This confirms that we implemented the \cite{li2024} method correctly, and that the halo sampling and bootstrap errors recover the analytic fraction.}
    \label{fig:ffb_fraction}
\end{figure}
\begin{figure*}
    \centering
    \includegraphics[height=0.92\textheight]{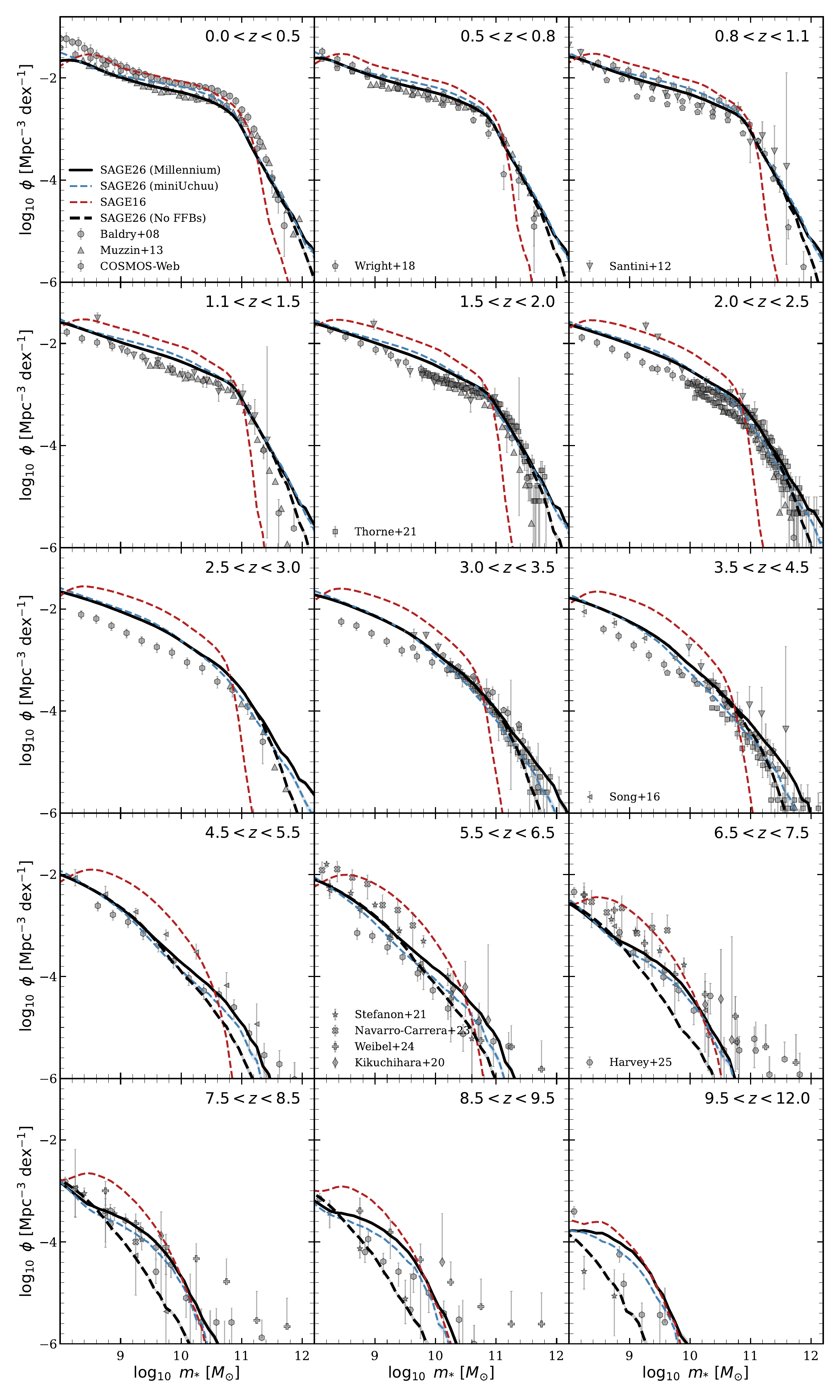}
    \caption{Galaxy stellar mass function across 15 redshift bins from $\rm z=0$ to $\rm z=12$, comparing SAGE26 Millennium (black solid line) and miniUchuu haloes (blue dashed line), SAGE26 without FFB galaxies (black dashed line) and SAGE16 (red dashed line). SAGE26 shows good agreement with observations across all redshifts. Two complementary effects drive the improvement at $\rm z>2$: the low-mass end is suppressed by $0.3-1.0$ dex relative to SAGE16, and the high-mass end is enhanced by $0.1-0.8$ dex relative to SAGE16. We also show observations from \citet{baldry2008,santini2012,muzzin2013,song2016,wright2018,kikuchihara2020,thorne2021,stefanon2021,weaver2023,navarro-carrera2024,shuntov2025,weibel2024} and \citet{harvey2025} for comparison.}
    \label{fig:smf_redshift_grid}
\end{figure*}
Recent JWST observations have revealed an unexpectedly high abundance of massive galaxies at very-high redshifts ($\rm z\geq6$), challenging standard galaxy formation models that predict less efficient mass assembly at these early epochs \citep{yung2020,donnan2022,labbe2023,finkelstein2023,harikane2023,lu2024}.\\
\\
At ultra-high redshifts, the Universe has higher gas densities, more compact galaxies and shorter dynamical times. It has been proposed that such conditions promote episodes of highly efficient star formation that temporarily overwhelm the regulation from supernova feedback \citep{dekel2023, li2024}, so-called feedback-free bursts (FFBs). During these FFBs, galaxies convert their gas reservoirs into stars at efficiencies far exceeding those typical of lower redshifts, which \cite{dekel2023} claims more naturally resolves the model-observation high-redshift massive galaxy tension. In SAGE26, we adopt this FFB mode, and refer the reader to Bradley et al., in prep. for a full exploration of FFB galaxies.\\
\\
FFBs occur when star-forming gas clouds reach densities where their gravitational free-fall time and gas cooling time become shorter than the timescale required for massive stars to evolve and explode as supernovae. Because this process relies on extreme gas densities, it is sensitive to cosmic expansion. At lower redshifts, gas densities decline and dynamical time increases, making it increasingly difficult for the free-fall time to outpace the feedback clock.\\
\\
Following \cite{dekel2023}, the conditions for FFBs can be modelled with a $\rm M_{vir}-z$ threshold leading to an increase in star formation efficiency:
\begin{equation}
    \rm \frac{M_{vir,\ FFB}}{10^{10.8}\ M_{\odot}}= \left(\frac{1+z}{10}\right)^{-6.2}.
\label{eq:ffb_threshold}
\end{equation}
At $\rm z\sim9$ the FFB threshold mass is approximately $\rm M_{vir,\ FFB}=10^{10.8}\ M_{\odot}$, which becomes less demanding at higher redshifts dropping to approximately $\rm M_{vir,\ FFB}=10^{9.7}\ M_{\odot}$ at $\rm z=14$. Rather than applying a binary cut-off, we follow \cite{li2024} and model the shift in galaxy population using a smooth transition across the vicinity of the threshold line, taking the fraction of FFB galaxies (assigned stochastically and redrawn at each snapshot) to be:
\begin{equation}
    f_{\rm FFB}(\rm M_{vir},z)=\rm \mathit{S}\left(\frac{log[M_{vir}/M_{vir,\ FFB}(z)]}{\Delta_{log\ M}}\right),
    \label{eq:ffb}
\end{equation}
where we use a sigmoid function, $\rm \mathit{S}(x)=(1+e^{-x})^{-1}$ with $\rm \Delta_{log\ M}$ chosen to be 0.15 dex \citep{li2024}. The smooth sigmoidal transition captures the shift in galaxy populations as they cross into the regime where higher gas densities enable feedback-free burst behaviour.\\
\\
We present the resulting behaviour (lines for the \citealt{li2024} model and markers for the SAGE26 haloes binned by mass) in \Fig{fig:ffb_fraction} as a function of halo mass. SAGE26 reproduces the analytic $f_{\rm FFB}$ of \cite{li2024} by construction. \Fig{fig:ffb_fraction} checks whether we have implemented the method correctly and are able to reproduce the analytic behaviour of \cite{li2024} from halo sampling and bootstrap errors. Due to the strong redshift evolution of the FFB threshold, $\rm M_{vir,FFB}\propto(1+z)^{-6.2}$, the transition from non-FFB haloes to FFB haloes occurs at progressively lower halo masses at higher redshifts.\\
\\
When a galaxy is in the FFB mode, we modify its star formation rate to be:
\begin{equation}\label{eq:FFB}
    \dot{\rm m}_{*}=\rm \alpha_{eff}\ \frac{m_{ISM}}{t_{dyn,\ disk}},
\end{equation}
where $\rm m_{ISM}$ is the interstellar medium or total cold gas reservoir. FFB galaxies differ from the standard prescription by forming stars at an increased efficiency, $\rm \alpha_{eff}=0.2$ rather than $\rm \alpha_{SF}=0.05$, and drawing on the total cold gas rather than the molecular hydrogen component.\\
\\
The inclusion of FFB galaxies improves the high-redshift stellar mass function and aligns the SAGE26 galaxy population with recent JWST observations, as seen in \Fig{fig:smf_redshift_grid}. Here we show the stellar mass function from $\rm z=0-12$, using SAGE26 run on Millennium and miniUchuu haloes, and observations from \cite{baldry2008,santini2012,muzzin2013,song2016,wright2018,kikuchihara2020,thorne2021,stefanon2021,weaver2023,navarro-carrera2024,shuntov2025,weibel2024} and \cite{harvey2025}. We also show the SAGE26 model without the feedback-free burst galaxies. SAGE26 shows significant improvement over its previous iteration and matches observations for most of cosmic history.\\
\\
For redshifts $\rm z=0-2$, SAGE26 can reproduce the low-mass end of the stellar mass function with a stable \enquote{knee} at $\rm 10^{10.5-11}\ M_{\odot}$, maintaining agreement with observations through to $\rm z=12$, within observational limits. The improved low-mass galaxy abundances at $\rm z>2$ are due to the modified stellar feedback model from FIRE (\Sec{SN_method}); this behaviour now properly suppresses the low-mass galaxy star formation, as supernovae can now reheat and eject a substantial amount of gas, in comparison to SAGE16.\\ 
\\
The new high-redshift behaviour in SAGE26 brings the model into much better agreement with observational constraints. The observational data points (grey) at $\rm z>4$ consistently fall below the SAGE16 predictions but align well with SAGE26 results. Crucially, SAGE26 tracks closely with the recent COSMOS-Web 2025 measurements \citep{shuntov2025}, which indicate elevated but not extreme star formation rates at high redshift.\\
\\
SAGE26 is one of several attempts to reconcile galaxy formation models with the abundance of galaxies at cosmic dawn. \cite{somerville2025} modulates star formation efficiency by gas surface density, following cloud-scale simulations, and reproduces the observed number densities within a semi-analytic model. \cite{lagos2026} argue that starbursts triggered by violent disk instabilities raise the variance of ultraviolet luminosity at fixed halo mass sufficiently to match the luminosity functions at high redshifts. \cite{boylan-kolchin2025} provides a physical pathway for dark matter accelerated star formation resulting in a mechanism for highly efficient galaxy growth during the early Universe. The fact that several different mechanisms can reproduce the same observables indicates the present constraints do not yet discriminate between them, and further routes are likely to be proposed as the data improves. See Bradley et al., in prep. for further exploration of these points.
\section{Discussion}\label{Discussion}
\begin{figure*}
    \centering
    \includegraphics[width=\linewidth]{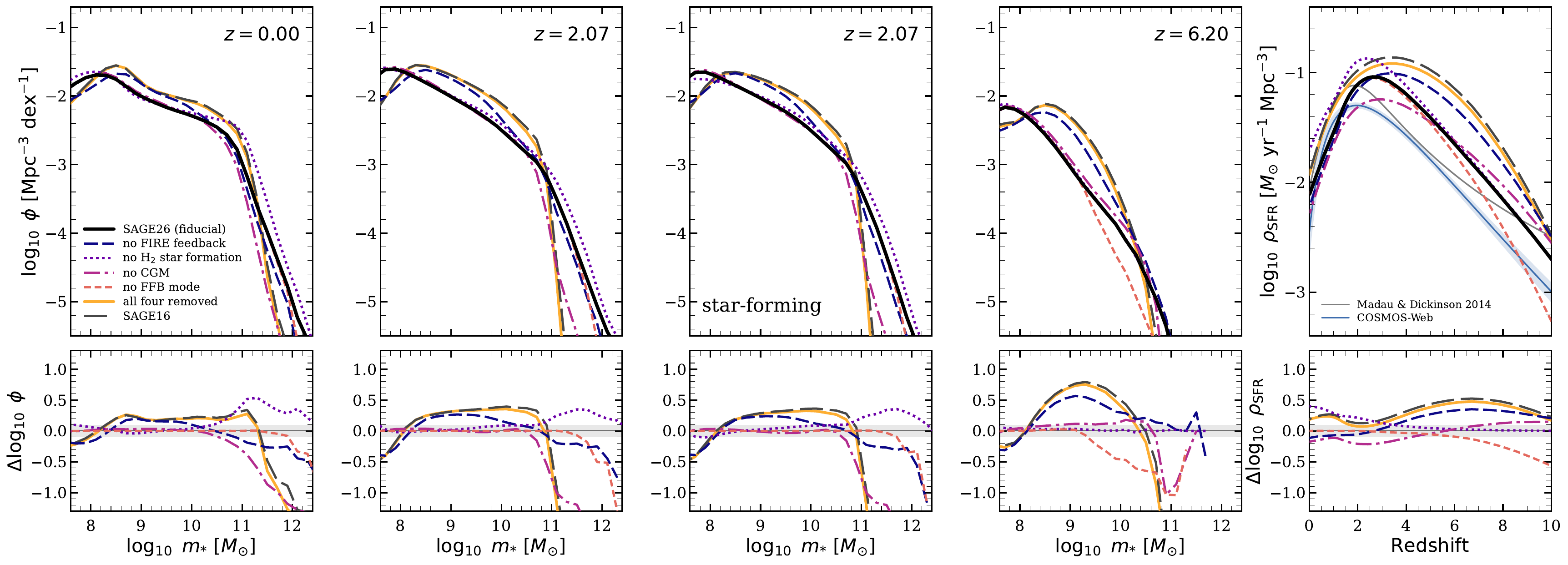}
    \caption{The contribution of each new ingredient. The columns show the stellar mass function at $\rm z=0,\ 2.07\ and\ 6.20$, the star-forming stellar mass function at $\rm z=2.07$, and the cosmic star formation rate density evolution, with the offset from the fiducial model (black) beneath each. Coloured lines disable one physics switch with nothing re-calibrated; a positive offset means that removing that ingredient raises the abundance, and yellow disables all four. The SAGE16 (grey) model was uncalibrated, and shading marks offsets below 0.1 dex.}
    \label{fig:ablation}
\end{figure*}
\subsection{New Prescription Interactions}\label{interactions}
SAGE26's new prescriptions work together to regulate the baryon cycle in a physically coherent way across cosmic time: CGM inflow provides the fuel required for FFB galaxies to form efficiently, FIRE-motivated supernova feedback regulates their growth, and molecular hydrogen regulates gas availability. To test this, we ran the model with each ingredient disabled in turn and once with all four disabled together, and no re-tuning between runs. In \Fig{fig:ablation} we show the outcome of this experiment using the stellar mass function at three redshifts and the star formation rate density evolution.\\
\\
Removing the FIRE feedback raises the abundance of $\rm 10^{8.5-9.5}\ M_{\odot}$ galaxies by $\rm 0.21-0.25$ dex at $\rm z=2$, and by $\rm 0.34-0.51$ dex at $\rm z=6$. By $\rm z=0$, this contribution decreases to approximately $\rm 0.12-0.15$ dex. The updated supernova feedback does most of its work early, suppressing dwarf galaxy abundances by a factor of $2-3$ at $\rm z=6$ but only 30\%-40\% by the present day, as expected from the $\rm (1+z)^{1.25}$ scaling in \Eq{eq:etareheat}.\\
\\
Disabling the molecular hydrogen prescription changes the low-mass abundance by 0.03 dex or less but raises the $\rm z=0$ quiescent fraction at $\rm m_*=10^{8.5}\ M_{\odot}$ from 27\% to 52\%. The molecular hydrogen prescription does not set how many dwarf galaxies form, but whether they are still forming stars, and it converts a population that would otherwise be quiescent into a star-forming one, without changing the mass function.\\
\\
The new CGM mechanics dominate at the massive end of the stellar mass function; removing them lowers the abundance of $\rm 10^{11.5}\ M_{\odot}$ galaxies at $\rm z=0$ by 0.33 dex and at $\rm z=2$ 0.44 dex, and reduces the stellar mass density by 17\%. Below $\rm M_{shock}$, the CGM reservoir drains at $\rm t_{cool,eff}+t_{ff}$, so free-fall sets the floor on how fast gas can cool in these haloes, while above $\rm M_{shock}$, cold streams are allowed to supply more fuel to the galaxies and increase the abundances in the stellar mass function.\\
\\
The FFB model is confined to high redshifts; removing it lowers the abundance at $\rm z=6$ of $\rm 10^{9.5-10.5}\ M_{\odot}$ galaxies by $\rm 0.21-0.65$ dex and the star formation density at $\rm z=8$ by 0.28 dex. The FFB mode is what lifts the massive end at $\rm z=6$ to the observed abundances, contributing up to a factor of 4.5 while leaving the model unchanged at later epochs.\\ 
\\
Summing the four individual offsets and comparing them with the joint run measures their interaction. The residual is small at low redshifts, 0.13 dex or less. For $\rm 10^{10.5}\ M_{\odot}$ galaxies at $\rm z=2$, the four taken individually predict a 0.66 dex decrease in abundance, while removing them together produces a 1.58 dex decrease. The star formation rate density behaves in a similar manner at the same epoch, and the residual grows to 0.23 dex for $\rm m_*=10^{9.5}\ M_{\odot}$ at $\rm z=6$. Each prescription dominates a different part of the mass-redshift plane, and none is redundant. The offsets add linearly where those regimes do not overlap and depart from linearity when they do, so the improvements are not independent corrections to SAGE16 but a coupled set.

\subsection{Calibration and Parameter Degeneracies}\label{calibration}

To manage the high-dimensional parameter space and avoid overfitting, we employed the SAGEswarm particle swarm optimisation package using a phased, incremental calibration strategy. For the sake of runtime speed, we used the mini-Millennium simulation, which has the same cosmology and particle mass as the Millennium simulation but a volume of $\rm (62.5\ h^{-1}\ Mpc)^3$. Initially, we focused exclusively on the baryonic cycle by calibrating the star formation and stellar feedback parameters, $\rm \alpha_{SF},\ \epsilon_{disk}\ and\ \epsilon_{halo}$. This group of parameters was constrained by matching the $\rm z=0$ stellar mass function and $\rm H_2$ mass function. Once a stable baseline for galaxy growth was established, we introduced the AGN parameters, $\rm \kappa_{R},\ \kappa_{Q}\ and\ \mathit{f}_{BH}$, constraining black hole growth and feedback against the local black hole-bulge mass relation.\\
\\
At each stage we minimise a weighted sum of the reduced $\chi^2$ over all constraints, adding observational and model uncertainties in quadrature, $\sigma_i^2=\sigma_{{\rm obs},i}^2+\sigma_{{\rm mod},i}^2$, where $\sigma_{\rm mod}$ is the Poisson uncertainty on the number of model galaxies per bin:
\begin{equation}
    \rm F=\sum_c \frac{w_c}{\sum_c w_c}\,\frac{1}{N_c}\sum_{i=1}^{N_c} \left(\frac{y_{{\rm mod},i}-y_{{\rm obs},i}}{\sigma_i}\right)^2 ,
    \label{eq:objective}
\end{equation}
with $\rm N_c$ the number of data points in constraint c and $\rm w_c=1$ throughout. \Tab{tab:calibration} lists the achieved value of F on each constraint for both SAGE26 and SAGE16.\\
\begin{table}
    \centering
    \begin{tabular}{lllrccc}
    \hline
    Constraint & & & & $N_{\rm obs}$ & SAGE16 & SAGE26 \\
    \hline
    SMF, $z=0$      & & & &  32 &   0.58 &  2.67 \\
    SMF, $z=0.5$    & & & &  38 &  77.96 &  12.53 \\
    SMF, $z=1$      & & & &  52 & 171.33 & 43.65 \\
    H$_2$MF, $z=0$  & & & &  11 & \nodata &  1.99 \\
    BHBM, $z=0$     & & & &   6 &  18.78 &  8.24 \\
    CSFRD          & & & &  25 &  66.97 & 17.10 \\
    \hline
    \hline
    All             & & & & 164 &  67.25 & 14.36 \\
    \hline
    \end{tabular}
    \caption{Achieved value of the calibration objective, \Eq{eq:objective}, for SAGE16 and SAGE26 on each constraint (SMF and $\rm H_2$MF are the stellar and molecular hydrogen mass functions, respectively; BHBM is the black hole to bulge mass relation, and the CSFRD is the cosmic star formation rate density), evaluated on mini-Millennium against the observational compilation of \Fig{fig:smf_redshift_grid}. Values are the reduced $\chi^2$ per observational data point; $\rm N_{obs}$ counts data points, not galaxies. SAGE16 does not separate the cold gas into atomic and molecular phases and so makes no prediction for the H$_2$ mass function. SAGE26 improves the total by a factor of $4.6$.}
    \label{tab:calibration}
\end{table}\\
Holding all but one parameter at its adopted value and sweeping the remaining one across its full range raises the objective function by between $4-76$ units for every parameter except $\rm \alpha_{eff}$. We then incrementally added higher-redshift constraints, specifically extending the stellar mass function to $\rm z=1$ and introducing the cosmic star formation rate density to $\rm z=4$. This joint, multi-epoch calibration ensures that the model captures genuine physical evolution across cosmic time, rather than artificially tuning parameters to compensate for localised errors.\\
\\
Across most tested redshifts, SAGE26 performs better than SAGE16, and the margin widens with redshift; at $\rm z=0.5$, the objective function for the stellar mass function falls from 77.96 for SAGE16 to 12.53 for SAGE26, and at $\rm z=1$, SAGE16 scores 171.33 while SAGE26 scores 43.65. This is reinforced by the cosmic star formation rate density, where SAGE16 scores 66.97 and SAGE26 scores 17.10 out to $\rm z=4$; SAGE26 improves over SAGE16 by a factor of $\sim 4.6$.\\
\\
While no single parameter improves the objective function when varied alone, an independent swarm run reaches an optimum faster by moving several parameters together, so the objective defines a degenerate valley rather than a unique minimum. The same degeneracy applies to SAGE16, where independent swarm runs reach comparable values at different parameter combinations, often by driving both supernova and AGN feedback parameters past the limits of their physical ranges. We therefore retain values close to SAGE16 for the six free parameters the two models share, so comparisons throughout this paper isolate the new physics rather than a change in calibration. The adopted set should be read as one point in that valley, chosen for continuity rather than a global optimum.\\
\\
Within this joint calibration, the supernova feedback parameters require careful attention to balance competing constraints. The normalisation $\rm \epsilon_{disk}=2.9$ for the mass-loading follows the functional form established in \cite{muratov2015}, directly from the FIRE simulations. We note that this specific value represents our calibration to match the observed stellar mass function across different epochs. Similarly, our adoption of $\rm \epsilon_{halo}=0.3$ for supernova-driven ejection aligns with the SAGE16 value, chosen to prevent excessive hot gas accumulation in intermediate-mass haloes at $\rm z=2-4$.\\
\\
For the feedback-free burst galaxies, we adopt $\rm \alpha_{eff}=0.2$ and $\rm \Delta logM=0.15$ following \cite{li2024}, chosen to match the observed stellar mass functions for $\rm z\approx6-8$ without over-producing galaxies at $\rm z>10$. Alternative values, $\rm 0.2\leq\alpha_{eff}\leq1.0$, are also plausible \citep{fu2026a,li2024}, suggesting a degree of degeneracy that future work should explore. The steep redshift evolution of $\rm M_{vir,\ FFB}\propto(1+z)^{-6.2}$ reflects the rapidly declining gas densities and increasing dynamical times as the Universe expands \citep{li2024}. This steep redshift evolution is not a free parameter, but is derived from the condition that the free-fall timescale must remain shorter than the feedback timescales.

\subsection{Physical Assumptions and Limitations}
Several deliberate simplifications in SAGE26 are worth examining, as they define the model's boundaries and naturally point to future work and improvements.\\
\\
The radio and quasar AGN feedback prescriptions are inherited from SAGE16, which remain unchanged in SAGE26. These modes reliably reproduce the black hole-bulge mass relation (\Fig{fig:BHBR}) and the basic quenching behaviour (\Fig{fig:sf_q_smf}), consistent with observations. However, we do not incorporate recent insights from X-ray observations and simulations of jet-gas interactions, which suggest a more complex interplay between AGN heating, cold flows and the circumgalactic medium (see \citealt{raouf2017}).\\
\\
For the CGM or hot halo assignment, varying $\rm M_{shock}$ by a factor of two in either direction with no calibration change leaves the stellar mass function unchanged to better than 0.02 dex below $\rm m_*=10^{10.5}\ M_{\odot}$ at every redshift and the cosmic star formation rate density unchanged to better than 0.04 dex. The massive end responds more strongly, by up to 0.29 dex at $\rm z=0$ and 0.37 dex at $\rm z=2$ above $\rm m_*=10^{11}\ M_{\odot}$, with a lower threshold suppressing massive galaxies as haloes enter the hot regime earlier and a higher threshold enhancing them.\\
\\
If the threshold were imprinted on the stellar mass function, it would appear near $\rm m_*\approx10^{10}\ M_{\odot}$, the median stellar mass of centrals within a factor of two of $\rm M_{shock}$. No such feature appears because scatter in stellar mass at fixed halo mass smooths the transition. The largest curvature lies at the exponential cut-off, and the spread in stellar mass at fixed halo mass smears the transition out at the population level. The abundance of massive galaxies at cosmic noon is therefore the measurement most sensitive to this parameter, as expected, since it determines whether such haloes accrete cold gas or cool radiatively.\\
\\
The transition dominates reproducing the correct number densities of massive galaxies at cosmic noon, while quenching occurs at later times. However, this model does not account for stream disruption by AGN-driven bubbles or magnetic fields, which could significantly affect massive systems. The consequences of this choice are confined to the highest-mass haloes at intermediate redshifts, where the cold stream contribution is subdominant.\\
\\
In addition, the galactic radii prescriptions, while improved with bulge-size tracking, remain simplified. For more detailed morphological studies, SAGE26 will require prescriptions that incorporate angular momentum and feedback-driven expansion.\\
\\
The FFB implementation (see \Sec{FFBs_method}) adopts the halo mass-redshift threshold of \cite{dekel2023} and \cite{li2024} rather than deriving it within SAGE26 by comparing timescales directly, as the model tracks cooling times natively. Deriving the criterion self-consistently, so that the burst phase is triggered by the model's own cloud-scale conditions rather than by an externally imposed threshold, is left to future work, to which we refer the reader to Bradley et al., in prep. for a fuller discussion.

\subsection{Numerical Limitations and Observational Uncertainties}
All of our results at much higher redshifts ($\rm z>8$) are affected by the Millennium and miniUchuu simulation mass resolution and snapshot cadence, where snapshots are few and far between. We note that high-redshift stellar mass estimates from JWST remain subject to systematic uncertainties arising from SED-fitting assumptions and dust models \citep{harvey2025}.\\
\\
The FFB threshold falls below 100 particle haloes above $\rm z\approx8$ in Millennium and $\rm z\approx9.4$ in miniUchuu, so at higher redshifts every resolved halo already lies above it, even though the massive galaxies in our high-redshift results are well resolved. Snapshot cadence is likewise coarse at these epochs, with nine Millennium snapshots between $\rm z=10-20$ and miniUchuu having four between $\rm z=10-14$.\\
\\
The two simulations differ in cosmology, mass resolution, snapshot cadence and tree builder, yet their SAGE26 SMFs agree closely at $\rm z=0$. The mean offset is $+0.01$ dex below $\rm 10^{10}\ M_{\odot}$ and $+0.05$ dex above it, with a largest single-bin difference of 0.18 dex. A resolution difference would grow towards low stellar mass, whereas the offset here is quite flat, so the residuals that are evident reflect the different cosmologies and volumes rather than resolution. These results indicate that the inconsistencies between the Millennium and miniUchuu simulations when run with SAGE16 that motivated aspects of this work are mostly resolved.\\
\\
Our improvements span multiple independent observables (the stellar mass function, star formation rate density and others) across a wide redshift range. The mass-metallicity relation, outflow rates, bulge sizes, and the $\rm z>1$ stellar mass functions are reproduced without being calibrated against, suggesting that the model captures genuine physical evolution rather than fitting systematic errors.\\
\\
Future application of SAGE26 to higher-resolution simulations will enable a more detailed exploration of dwarf galaxy populations and the physics of satellite quenching. Using the Shin-Uchuu simulation \citep{ishiyama2021} would be the next natural step, as it provides a substantially higher particle mass resolution of $\rm 8.97\times10^5\ h^{-1}\ M_{\odot}$. Such higher-resolution simulations enable the exploration of satellite-galaxy physics and smaller-scale phenomena in the galactic fountain.\\
\\
The circumgalactic medium framework developed here provides a foundation for incorporating environmental effects on gas accretion and stripping in group and cluster environments. In particular, SAGE26 now tracks the formation and assembly channels of intracluster stars, as well as full star formation histories, enabling a detailed study of their formation pathways and their connection to halo assembly history. This will be the subject of upcoming work.
\section{Conclusion}
We present SAGE26, a comprehensive modernisation of the Semi-Analytic Galaxy Evolution model that addresses systematic discrepancies between theoretical predictions and observations over cosmic time. SAGE26 integrates five key physics improvements:
\begin{itemize}
    \item Distinct circumgalactic medium and hot-halo gas physics (\Fig{fig:SAGE26_schematic}) with a critical transition at $\rm M_{shock}=6\times10^{11}\ M_{\odot}$ (\Sec{CGM_method}).
    \item Molecular hydrogen formation that suppresses star formation in low-surface-density dwarf galaxies, improves the quiescent galaxy population (\Fig{fig:sf_q_smf}), and links gas density to stellar production (\Sec{h2}).
    \item Velocity- and redshift-dependent supernova feedback (\Fig{fig:massloading}) based on the FIRE simulations (\Sec{SN_method}), improving regulation of low- to intermediate-mass galaxy growth (\Fig{fig:smf_redshift_grid}).
    \item Observationally based bulge radii tracking through time (\Sec{bulges}) that can differentiate between merger- and disk-instability-driven formation events (\Fig{fig:bulgetype}). 
    \item Feedback-free burst galaxy growth at high redshifts (\Sec{FFBs_method}) where star formation operates at an enhanced efficiency ($\rm \alpha_{eff}=0.2$).
\end{itemize}
These new prescriptions work to regulate galaxy formation from cosmic dawn to the present day.\\
\\
We find that our SAGE26 model achieves improvements across multiple observational benchmarks. The stellar mass function now matches observations with a median absolute offset of 0.20 dex from $\rm z=0-12$, rising from 0.13 dex below $\rm z=1$ to $\rm 0.2-0.5$ dex above $\rm z=6$, where the observations differ from themselves as much as the model does. The cosmic star formation rate density now declines by a factor of $\sim3$ more steeply than SAGE16 at $\rm z=6$ and tracks COSMOS-Web and other observational constraints.\\
\\
With SAGE26, we substantially reduce the excess of quiescent dwarfs produced by SAGE16 and increase the number density of high-mass quiescent galaxies. With a new model for the circumgalactic medium, along with new physics for pristine infalling gas, we now enable a new pathway for fuelling star-forming galaxies. Most significantly, the feedback-free burst implementation improves high-redshift predictions, increasing the abundance of massive galaxies by $0.21-0.65$ dex at $\rm z>6$ to match recent JWST observations while maintaining excellent agreement at lower redshifts.\\
\\
SAGE26 demonstrates that semi-analytic models remain powerful tools for understanding galaxy formation when grounded in modern physical insights from observations and simulations. The model maintains SAGE16's computational efficiency, running on most N-body simulations, and offering physically transparent prescriptions that enable clear interpretations of the drivers of galaxy evolution. SAGE26 is publicly available and part of a suite\footnote[2]{\url{https://github.com/sage-home}} that includes SAGE16, SAGEswarm (parameter calibration), LightSAGE (light cone generation) and ViSAGE (data visualisation). As JWST, Euclid, and next-generation telescopes and surveys continue to probe the local Universe through cosmic dawn and beyond, SAGE26 provides both a framework for interpreting these discoveries and a platform for predicting what remains to be discovered in the Universe.
\begin{acknowledgments}
DJC is a recipient of an Australian Research Council Future Fellowship (project No. FT220100841) funded by the Australian Government.\\
\\
DJC, RAMP, JW and MB acknowledge this research was supported by the Australian Government through the Australian Research Council Centre of Excellence for Dark Matter Particle Physics (CDM, CE200100008).\\
\\
This work used the SCIPY \citep{virtanen2020}, NUMPY \citep{harris2020}, MATPLOTLIB \citep{hunter2007}, PANDAS \citep{mckinney2010}, ASTROPY \citep{theastropycollaboration2022}, and COLOSSUS \citep{diemer2018} packages for PYTHON.\\
\\
We used the OzSTAR supercomputer at Swinburne University of Technology for many aspects of this work, including running SAGE26. The OzSTAR supercomputing environment is managed through the Centre for Astrophysics \& Supercomputing with support from Swinburne IT. The supercomputing program receives continued financial support for operations from Astronomy Australia Limited and the Australian Commonwealth Government through the National Collaborative Research Infrastructure Strategy (NCRIS). The hardware for Ngarrgu Tindebeek was purchased through a grant from the Victorian Higher Education State Investment Fund (VHESIF).\\
\\
We thank C. Lagos and G. Kacprzak for their contributions and guidance.
\end{acknowledgments}
\section*{Data Availability}
SAGE26 is publicly available at \url{https://github.com/sage-home/SAGE26}, SAGEswarm is available at \url{https://github.com/sage-home/SAGEswarm}, and the interactive ViSAGE tool is available at \url{https://github.com/sage-home/ViSAGE}.\\
\\
The Millennium Simulation was carried out by the Virgo Supercomputing Consortium at the Computing Centre of the Max Planck Society in Garching. It is publicly available at \url{https://www.mpa-garching.mpg.de/Millennium/}. The Uchuu Simulation was carried out by the Uchuu Collaboration and is publicly available at \url{https://www.skiesanduniverses.org/Simulations/Uchuu/}.

\bibliographystyle{aasjournalv7}
\bibliography{SAGE_bib}{}



\end{document}